\documentclass[12pt]{iopart}

\usepackage{iopams}
\usepackage{graphicx}
\usepackage{dcolumn}
\usepackage{bm}
\eqnobysec

\begin{document}

\title[Cosmology based on relativity with a privileged frame]{Cosmological models based on relativity with a privileged frame}

\author{Georgy I. Burde}

\address{Swiss Institute for Dryland Environmental and Energy Research,\\
Jacob Blaustein Institutes for Desert Research,
Ben-Gurion University of the Negev,\\
Sede-Boker Campus, 84990, Israel}
\ead{georg@bgu.ac.il}
\vspace{10pt}

\begin{abstract}
Special relativity (SR) with a privileged frame is a framework, which, like the standard relativity theory, is based on the relativity principle and the universality of the (two-way) speed of light but includes a privileged frame as an essential element.
It is developed using the following first principles: (1) Anisotropy of the one-way speed of light in an inertial frame is due to its motion with respect to the privileged frame; (2) Space-time transformations between inertial frames leave the equation of anisotropic light propagation
invariant; (3) A set of the transformations possesses a group structure. The Lie group theory apparatus is applied to define groups of transformations.
The correspondingly modified general relativity (GR), like the standard GR, is based on
the equivalence principle but with the properly modified space-time local symmetry in which an invariant combination differs from the Minkowski interval of the standard SR. That combination can be converted into the Minkowski interval by a change of space-time variables and then the complete apparatus of general relativity can be applied in the new variables. However, to calculate physical effects, an inverse transformation to the 'physical' time and space intervals is to be used.
Applying the modified GR to cosmology yields the luminosity distance -- redshift relation corrected such that the observed deceleration parameter can be negative as it was derived from the data for type Ia supernovae. Thus,
the observed negative values of the deceleration parameter can be explained within the matter-dominated Friedman-Robertson-Walker (FRW) cosmological model of the universe without introducing the dark energy.
A number of other observations, such as Baryon Acoustic Oscillations (BAO) and Cosmic Microwave Background (CMB),
that are commonly considered as supporting the late-time cosmic acceleration and the existence of dark energy, also can be well fit to the cosmological model arising from the GR based on the SR with a privileged frame.
\end{abstract}




\section{Introduction}

Special relativity (SR) underpins nearly all of present day physics.
The space-time symmetry of Lorentz invariance is one of the cornerstones of general relativity (GR) and other theories of fundamental physics.
Nevertheless, the modern view is that, at least cosmologically, a privileged reference frame does exist.
Modern cosmological models
are based on the assumption of existence of a privileged frame in which the universe appears isotropic to a "typical" freely falling observer. That typical (privileged) Lorentz frame is usually assumed to
coincide with the frame in which the cosmic microwave background (CMB) temperature distribution is isotropic ('CMB frame').

The view, that there exists a privileged frame of reference, seems to unambiguously lead to the abolishment of the basic principles of the special relativity theory: the principle of relativity
and the principle of universality of the speed of light. Correspondingly, the modern
versions of experimental tests of special relativity and the "test theories" of special relativity
\cite{Rob}, \cite{MS1}
presume that a privileged inertial reference frame, identified with the CMB frame, is the only frame in which
the \textit{two-way} speed of light (the average speed from source to observer and back) is isotropic
while it is anisotropic in relatively moving frames.
Furthermore, it seems that accepting the existence of a privileged frame forces one to abandon the group structure for the set of space-time transformations between
inertial frames -- in the test theories, transformations between "moving" frames are not considered, only a form of   transformations between a privileged  "rest" frame and moving frames is postulated.

The initial motivation for this study was investigate fundamentals of relativity by developing a theory which incorporates the privileged frame into the framework of SR while retaining the basic principles of the theory, the relativity principle and universality of the speed of light, and also preserving the group structure of the set of transformations between inertial frames.
However, after developing the theory that satisfies all those requirements, it was found that the special relativity with a privileged frame allows a straightforward extension to general relativity. Further, applying the modified general relativity to cosmology yields the luminosity distance versus redshift relation which provides an interpretation of the  type Ia supernovae data differing from the common one. That relation allows negative values of the deceleration parameter in the matter-dominated Friedman-Robertson-Walker cosmological model of the universe and so it does not obligatory require introducing a dark energy. Considering those cosmological applications
of the relativity with a privileged frame is the primary goal of this paper.

The main body of the paper consists of three parts. The first part is devoted to developing
the special relativity, which, like the standard relativity theory, is based on the relativity principle and the universality of the (two-way) speed of light, but includes a privileged frame as an essential element.
It is shown that the reconciliation and synthesis of those seemingly incompatible concepts is possible
in the framework of the relativity theory. Since any \textit{one-way}  speeds of light, consistent with the two-way speed equal to $c$, are acceptable, a privileged frame can be defined as the only frame in which the one-way speed of light is isotropic
while it is anisotropic in any other frame moving with respect to a privileged frame.
The analysis is based on invariance of the equation of anisotropic light propagation (preserving the property that the two-way speed of light is equal to $c$) with respect to the space-time transformations between inertial frames with the requirement  that a set of the transformations possesses a group structure. The anisotropy parameter in the equation of light propagation is treated as a variable which takes part in the group transformations varying from frame to frame. In such a framework, the principle of constancy and universality of the two-way speed of light and the group property are preserved. The principle of relativity is also preserved since the privileged frame, in which the anisotropy parameter is zero, enters the analysis on equal footing with other frames -- the transformations from/to that frame are  not distinguished from other members of the group of transformations. However, the existence of a privileged frame is an essential element of the framework since an argument, that a size of the anisotropy in a specific frame is determined by its velocity with respect to the privileged frame, is used to specify the transformations. 

At the first sight, that argument seems to be in conflict with a common view that, because of the inescapable entanglement between remote clock synchronization and one-way speed of light, the one-way speed of light is irreducibly conventional (see, e.g., \cite{Ung1}-- \cite{Rizzi}). Nevertheless, the present paper analysis demonstrates that, in an anisotropic system,
a specific value of the one-way speed of light (and the corresponding synchronization) is selected  in some objective way as a measure of anisotropy -- in the present context, it is
the anisotropy caused by motion of a system relative to the privileged frame.

The space-time transformations between inertial frames derived as a result of the analysis differ from the Lorentz transformations. Correspondingly, the interval between two events, as distinct from the standard SR, is not invariant under the transformations but conformally transformed. In other terms, a combination, which is invariant under the transformations (a counterpart of the interval of the standard SR),  differs from the Minkowski interval. In view of the fact that the theory is based on the special relativity principles, it means that the Lorentz invariance is violated without violation of relativistic invariance.

Since the local Lorentz invariance is one of the foundations of general relativity, the corresponding alterations need to be introduced into the framework of GR. The second part of the analysis is devoted to formulating a general relativity that is based on the equivalence principle but with modified space-time local symmetry
in which an invariant combination differs from the Minkowski interval of the standard SR. That combination can be converted into the Minkowski interval by a change of space-time variables and then the complete apparatus of general relativity can be applied in the new variables. However, to calculate physical effects, an inverse transformation to the 'physical' time and space intervals is to be used.

The third part of the paper is devoted to applying the modified GR to cosmological models which is a primary goal of this study. The cosmological models based on the modified GR allow an interpretation of the luminosity distance versus redshift relation for type Ia supernovae that is different from the common one. In the modern cosmology, that relation is interpreted as an indication that the present expansion of the universe is accelerated. This implies that the time evolution of the expansion rate cannot be described by a matter-dominated Friedman-Robertson-Walker cosmological model of the universe. In order to explain the discrepancy within the context of general relativity, dark energy,  a new component of the energy density with strongly negative pressure that makes the universe accelerate, is introduced. In the relativity with a privileged frame, the deceleration parameter in the luminosity distance -- redshift relation is corrected such that the deceleration parameter can be negative. Thus, in that framework, the observed negative values of the deceleration parameter can be explained within the Friedman model of the matter-dominated universe with no dark energy.

The late-time cosmic acceleration (and the existence of dark energy) is commonly considered to be supported by
a number of other observations such as Baryon acoustic oscillations (BAO) and CMB.
Nevertheless, the only observational data, that may be considered as providing a "direct" evidence for a dark energy, is the Hubble diagram of distant supernovae. As a matter of fact, what is usually shown is that the BAO and CMB measurements can be made consistent with the supernovae observations by specifying the cosmological and dark energy parameters.
The present's paper analysis shows that both the SNIa data and the BAO results can be well fit to the model arising from the modified GR that is based on the relativity with a privileged frame. The analysis cannot be straightforwardly extended to calculating the CMB effects but it can be shown that
the model is not contradictory with the data. In general, the cosmological model based on the relativity with a privileged frame can provide an alternative to the cosmology with a dark energy.



The paper is organized, as follows. In Section 2, following the Introduction, the special relativity with a privileged frame is constructed. In Section 3, an extension to the general relativity is considered. In Section 4, a cosmological model based on that extension is developed and fitting the observational data to the model is discussed. Concluding comments are furnished in  Section 5. In Appendix A, the modified general relativity is applied to the astrophysical problem of collapse of a dust-like sphere. In Appendix B, some auxiliary calculations are placed.

\section{Special relativity}

\subsection{Conceptual framework}


The issue of anisotropy of the one-way speed of light is traditionally placed into the context of conventionality of distant simultaneity and clock synchronization \cite{Ung1}-- \cite{Rizzi}.
Simultaneity at distant space points
of an inertial system is defined by a clock synchronization that
makes use of light signals.
Let a pulse of light is emitted from the master clock
and reflected off the remote clock. If   $t_0$ and  $t_R$ are respectively the times of emission and reception of the light pulse at the master clock and $t$ is the time of reflection of the pulse at the remote clock then the conventionality of simultaneity is a statement that one is
free to choose the time $t$ to be anywhere between $t_0$ and  $t_R$. This freedom may be parameterized
by a parameter $k_{\epsilon}$, as follows
\begin{equation}\label{In1}
t=t_0+\frac{1+k_{\epsilon}}{2}\left(t_R-t_0\right);\quad \left\vert k_{\epsilon} \right \vert <1
\end{equation}
Any choice of $k_{\epsilon}\neq 0$ corresponds
to assigning different one-way speeds of light signals in each
direction which must satisfy the condition that the average is
equal to $c$. Speed of light in each direction is
therefore\begin{equation}\label{In2} V_{\pm}=\frac{c}{1\pm
k_{\epsilon}}
\end{equation}
The "standard" (Einstein) synchronization entailing equal speeds in opposite directions corresponds to
$k_{\epsilon}= 0$.
If the described procedure is used for setting up throughout the
frame of a set of clocks using signals from some master clock
placed at the spatial origin, a difference in the standard and
nonstandard clock synchronization may be reduced to a change of
coordinates \cite{Ung1}-- \cite{Rizzi}
\begin{equation}\label{In3}
t=t^{(s)}+\frac{k_{\epsilon} x}{c},\quad x=x^{(s)}
\end{equation}
where $t^{(s)}=(t_0+t_R)/2$ is the time setting according to
Einstein (standard) synchronization procedure.

The analysis can be extended to the three dimensional case. If a
beam of light propagates (along straight lines) from a starting
point and through the reflection over suitable mirrors covers a
closed part the experimental fact is that the speed of light as
measured over closed part is always $c$ (\textit{Round-Trip Light
Principle}). In accordance with that experimental fact, if the speed of light is allowed to be
anisotropic it must depend on the direction of propagation  as \cite{And}, \cite{Min}
\begin{equation}\label{In4a}
V=\frac{c}{1+\mathbf{k_{\epsilon}}\mathbf{n}}=\frac{c}{1+k_{\epsilon}\cos
\theta_k}
\end{equation}
where $\mathbf{k_{\epsilon}}$ is a constant vector and $\theta_k$
is the angle between the direction of propagation $\mathbf{n}$ and
$\mathbf{k_{\epsilon}}$.
Similar to the
one-dimensional case, the law (\ref{In4a}) may be considered as a
result of the transformation from "standard" coordinatization of
the four-dimensional space-time manifold, with $k_{\epsilon}=0$,
to the "nonstandard" one with $k_{\epsilon}\neq 0$:
\begin{equation}\label{In5}
t=t^{(s)}+\frac{\mathbf{k_{\epsilon}}\mathbf{r}}{c},\quad
\mathbf{r}=\mathbf{r}^{(s)}
\end{equation}


The freedom in the choice of synchronization has been repeatedly used in the literature to derive
the transformations which are treated as replacing standard Lorentz
transformations of special relativity if anisotropic one-way light speeds with  $k_{\epsilon}\neq 0$
are assumed -- see, e.g., \cite{Edw} -- \cite{Ung2}.
The derivations of those transformations (in what follows, they will be called the "$\epsilon$-Lorentz transformations", the name is due to \cite{WinII}, \cite{Ung2}) are based on kinematic arguments and the requirement that, in the case of $k_{\epsilon}=0$, the relations of the special relativity theory in its standard formulation were valid. The $\epsilon$-Lorentz transformations can be equally obtained from the standard Lorentz transformations by
a change of coordinates (\ref{In3}). The fact, that there can exist a variety of "anisotropic" kinematics with different $k_{\epsilon}$, is usually considered as supporting the view that the one-way speed of light is irreducibly conventional.

The purpose of the following discussion is to demonstrate that, in the case of an anisotropic system, that view is incorrect so that
a specific value of the one-way speed of light (and the corresponding synchronization) is selected  in some objective way as a measure of anisotropy. In particular, it is shown that (1) the variety of kinematics corresponding to the $\epsilon$-Lorentz transformations, which are commonly considered as incorporating anisotropy, are in fact not applicable to an anisotropic system and (2) in the  case of an isotropic system,  the particular case of the transformations corresponding to the isotropic one-way speed of light and Einstein synchronization (standard Lorentz transformations) is privileged in some objective way.

The statement (1) is related to the issue of
\textit{invariance of the interval}. Invariance of the interval
is traditionally considered as an integral part of the physics of special relativity which is
used as a starting point for derivation of the space-time transformations between inertial frames. Nevertheless, invariance of the interval
is not a straightforward consequence of the basic principles of the theory. The two principles constituting the conceptual basis of the special relativity,  the \textit{principle of relativity}, which states the equivalence of all inertial frames as regards the formulation of the laws of physics, and \textit{universality of the speed of light} in inertial frames,
taken together
lead to the condition of \textit{invariance of the equation of light propagation} with respect to the coordinate transformations between inertial frames.
Thus, in general, not the invariance of the interval but invariance of the equation of light propagation should be a starting point for derivation of the transformations.
Therefore, in the textbooks (see, e.g., \cite{Pauli}, \cite{LL}),  the use of the interval invariance is usually preceded
by a proof of its validity based on
invariance of the equation of light propagation. However, those proofs are not valid if an anisotropy is present and the same arguments lead to the conclusion that, in the presence of
anisotropy, the interval is not invariant but modified by a conformal factor \cite{burde}.
The "$\epsilon$-Lorentz transformations", like the standard Lorentz transformations, leave the interval invariant
and therefore they are applicable only to an isotropic system.

The statement (2)
relies on the \textit{correspondence principle}.
The correspondence principle was taken by Niels Bohr as the guiding principle to discoveries in the old quantum theory. Since then it
was considered as a guideline for the selection of new theories in physical science.
In the context of special relativity, the correspondence principle
implies that Einstein's theory
of special relativity reduces to classical mechanics in the limit
of small velocities in comparison to the speed of light.
Being applied to the special relativity kinematics, the correspondence principle
requires that \textit{the transformations between inertial frames
should turn into the Galilean transformations in the limit of
small velocities}.
The "$\epsilon$-Lorentz transformations"  do not satisfy the correspondence principle unless $k_{\epsilon}=0$ \cite{burde} which means that the isotropic one-way speed of light and Einstein synchrony are selected in some objective way if no anisotropy is present in a physical system.

On the basis of the above discussion one can conclude that, in the case of an \textit{anisotropic} system, there exists a privileged value of the one-way speed of light selected by the size of the anisotropy. Thus,
a value of the one-way speed of light acquires meaning of a measure of  a really existing anisotropy -- in the present context, it is the anisotropy caused by motion of a system relative to the privileged frame.



In what follows, the special relativity kinematics applicable to an anisotropic system
is developed based on the first principles of special relativity but without refereeing to the relations of the standard relativity theory.
The principles constituting the conceptual basis of special relativity, the relativity principle, according to which physical laws should have the same forms in all inertial frames, and the universality of the speed of light in inertial frames,
lead to the requirement of
invariance of the equation of light propagation with respect to the coordinate transformations between inertial frames. In the present context, it should be invariance of the equation of propagation of light which incorporates the anisotropy of the one-way speed of light, with the law of variation of the speed with direction (\ref{In4a}).
The anisotropic
equation of light propagation incorporating the law (\ref{In4a}) has the form \cite{burde}
\begin{equation}\label{Sec3.1_Int}
ds^2=c^2 dt^2-2k c\; dtdx-(1-k^2)dx^2-dy^2-dz^2=0
\end{equation}
where $(x,y,z)$ are coordinates, $t$ is time and $\mathbf{k}$ is a (constant) vector characteristic of the anisotropy.
The change of notation, as compared with (\ref{In4a}), from
$k_{\epsilon}$ to $k$ is intended to indicate that $\mathbf{k}$ is
a parameter value corresponding to the size of the really existing anisotropy
while $k_{\epsilon}$ defines the anisotropy in the one-way speeds
of light due to the nonstandard synchrony equivalent to the coordinate change
(\ref{In5}).
Note that although the form (\ref{Sec3.1_Int}) is usually attributed to the
one-dimensional formulation,
in the three-dimensional case, the equation has the same form if
the anisotropy vector $\mathbf{k}$ is directed along the $x$-axis \cite{burde}.





Further, in the development of the anisotropic relativistic
kinematics, a number of other physical requirements,
associativity, reciprocity and so on are to be satisfied which all
are covered by the condition that the transformations between
the frames form a group.
Thus, the group property should be taken as another first
principle.
The formulation based on the invariance and group property suggests using the \textit{Lie group theory} apparatus for defining groups of space-time transformations between inertial frames.

At this point, it should be clarified that there can exist two different cases: (1)
The size of anisotropy does not depend on the observer motion and so is the same in all inertial frames; (2) The anisotropy is due to the observer motion with respect to a privileged frame and so
the size of anisotropy varies from frame to frame. Groups of space-time transformations for the first case are studied in \cite{burde}. The second case is relevant to the subject of the present study. In that case, the anisotropy parameter becomes a variable which takes part in the transformations so that groups of transformations in \textit{five} variables $\{x,y,z,t,k\}$ are to be studied.
In such a framework, the privileged frame,
commonly defined by that the propagation of light in that frame is isotropic, is
naturally present as the frame in which $k=0$. However, it does not violate the relativity principle since the transformations from/to that frame are  not distinguished from other members of the group. Nevertheless, the fact, that the anisotropy of the one-way speed of light in an arbitrary inertial frame is due to motion of that frame relative to the privileged frame,
is a part of the paradigm  which allows to specify the transformations.

The procedure of obtaining the transformations consists of the following steps:  (1) The infinitesimal invariance condition is applied to the equation of light propagation which  yields determining equations for the infinitesimal group generators; (2) The determining equations are solved to define the group generators and the \textit{correspondence principle} is applied to specify the solutions; (3) Having the group generators defined the finite transformations are determined as solutions of the Lie equations; (4) The group
parameter is related to physical parameters using some obvious conditions; (5) Finally, the conceptual argument, that the size of anisotropy of the one-way speed of light in an arbitrary inertial frame depends on its velocity relative to the privileged frame,
is used to specify the results and place them into the context of special relativity with a privileged frame

The transformations between inertial frames derived in
such a way contain a scale factor
and thus do not leave the interval
between two events invariant but modify it by a conformal factor
(square of the scale factor). Applying the conformal invariance in physical theories originates from the papers by Bateman \cite{Bate} and Cunningham \cite{Cunn} who discovered the form-invariance of Maxwell's equations for electromagnetism with respect to conformal space-time transformations. Since then conformal symmetries have been successfully exploited
for many physical systems (see, e.g., reviews
\cite{Fult}, \cite{Kast}).
Transformations which conformally modify
Minkowski metric have been introduced in the context of the special relativity kinematics in the presence of space anisotropy in \cite{Bogoslov77} and \cite{Sonego} (see also \cite{Lalan}). 
As a matter of fact, those works are not directly related to the subject of the present study as they consider the case of a constant anisotropy degree, not dependent on the frame motion. Nevertheless, it is worthwhile to note that in the works \cite{Bogoslov77}, \cite{Sonego}
the assumption that the form of the metric changes by a "conformal" factor is \textit{imposed} while, in the framework of the present analysis, conformal invariance of the metric \textit{arises} as an intrinsic feature of special relativity based on invariance of the anisotropic equation of light propagation and the group property (see \cite{burde} for a more detailed discussion of the papers \cite{Bogoslov77}, \cite{Sonego}).


\subsection{Space-time transformations with a varying anisotropy parameter}


Consider two arbitrary inertial reference frames $S$ and $S'$ in
the standard configuration with the $y$- and $z$-axes of the two
frames being parallel while the relative motion is along the
common $x$-axis. The space and time coordinates in $S$ and $S'$
are denoted respectively as $\{X,Y,Z,T\}$ and $\{x,y,z,t\}$. The
velocity of the $S'$ frame along the positive $x$ direction in
$S$, is denoted by $v$. It is assumed that the frame $S'$ moves relative to $S$ along
the direction determined by the vector $\mathbf{k}$. This assumption is justified by that one
of the frames in a set of frames with different values of $k$
is a privileged frame, in which $k=0$, so that the transformations must include, as a particular case, the transformation to that privileged frame. Since the anisotropy is attributed to the fact of motion with respect to the
privileged frame it is expected that the axis of
anisotropy is along the direction of motion (however, the direction of the anisotropy vector  can be both coinciding and opposite to that of velocity).

Transformations between the frames are
derived based on the following first principles: \textit{invariance of the equation of light
propagation} (underlined by the relativity principle), \textit{group property} and the \textit{correspondence principle}. Note that the group property is used not as in
the traditional analysis which commonly proceeds along the lines initiated by \cite{Ignatowski}
and \cite{FrRo} which are based on the linearity assumption and
relativity arguments. The difference can be seen from the derivation of the standard Lorentz transformations \cite{burde}.

\medskip

\noindent \textit{Invariance of the equation of light propagation}.
The equations for
light propagation in the frames $S$ and $S'$ are
\begin{eqnarray}\label{LPPref}
&& c^2 dT^2-2K c\; dTdX-(1-K^2)dX^2-dY^2-dZ^2=0,\\
&&\label{LPPref1}c^2 dt^2-2k c\;
dtdx-\left(1-k^2\right)dx^2-dy^2-dz^2=0
\end{eqnarray}
where the anisotropy parameters $K$ and $k$ in the frames $S$ and $S'$ are different. The relativity principle implies that the transformations of variables from $\{X,Y,Z,T,K\}$ to $\{x,y,z,t,k\}$ leave the form of the equation of light propagation invariant so that (\ref{LPPref}) is converted into (\ref{LPPref1}) under the transformations.

\medskip

\noindent\textit{Group property}.
The transformations between inertial frames form a one-parameter group
with the group parameter $a=a(v)$ (such that $v\ll 1$ corresponds
to $a\ll 1$):
\begin{equation}\label{2.1}
\eqalign{x=f(X,Y,Z,T,K;a), \quad y=g(X,Y,Z,T,K;a), \quad
z=h(X,Y,Z,T,K;a),\cr
t=q(X,Y,Z,T,K;a); \quad k=p(K;a)}
\end{equation}
Remark that $k$ is a transformed variable taking part in the group transformations.
Based on the symmetry arguments it is assumed that the
transformations of the variables $x$ and $t$ do not involve the
variables $y$ and $z$ and vice versa:
\begin{equation}\label{2.4K}
\eqalign{x=f(X,T,K;a), \; t=q(X,T,K;a),\quad y=g(Y,Z,K;a), \; z=h(Y,Z,K;a);\cr
k=p(K;a)}
\end{equation}

\medskip

\noindent \textit{Correspondence principle}.
The correspondence principle requires that, in the limit
of small velocities $v\ll c$ (small values of the group parameter
$a\ll 1$), the formula for transformation of the coordinate $x$
turns into that of the Galilean transformation\footnote{It should be noted that the relations $t=T$, $y=Y$ and $z=Z$, which are commonly included into the system of equations called the Galilean transformations, are not required to be valid in the limit of small velocities. Only the relation  (\ref{GalK}), which contains the first order term, provides a reliable basis for specifying the group transformations based on the correspondence principle (see more details in \cite{burde}).}
\begin{equation}\label{GalK}
x=X- v T
\end{equation}
Remark that the small $v$ limit is not influenced by the presence of anisotropy of the light propagation. It is evident that there should be no traces of light anisotropy in that limit, the issues of the light speed and its anisotropy are alien to the framework of Galilean kinematics.


The group property and the requirement of invariance of the
equation of light propagation suggest applying the infinitesimal Lie technique
 (see, e.g., \cite{Bluman}, \cite{Olver}).
The infinitesimal transformations corresponding to (\ref{2.4K}) are introduced, as follows
\begin{equation}\label{2.5K}
\eqalign{x\approx X+\xi(X,T,K)a, \quad t\approx T+\tau(X,T,K)a, \cr
y\approx Y+\eta(Y,Z,K)a, \quad z\approx Z+\zeta(Y,Z,K)a, \quad k \approx K+\kappa (K) a}
\end{equation}
The correspondence principle is applied to specify partially the infinitesimal group generators.
Equation (\ref{GalK}) is used to calculate the group generator
$\xi(X,T)$, as follows
\begin{equation}\label{2.6}
\xi=\left(\frac{\partial x}{\partial a}\right)_{a=0}=
\left(\frac{\partial \left(X-v(a) T\right)}{\partial
a}\right)_{a=0}=-v'(0) T
\end{equation}
It can be set $v'(0)=1$ without loss of generality since
this constant can be eliminated by redefining the group parameter. Thus, the generator $\xi$ is defined by
\begin{equation}\label{xiAnizK} \xi=-T
\end{equation}
Then equations (\ref{LPPref}) and (\ref{LPPref1}) are used to derive determining equations for
the group generators $\tau(X,T,K)$, $\xi(X,T,K)$, $\eta(Y,Z,K)$,
$\zeta(Y,Z,K)$ and $\kappa (K)$.
Substituting the infinitesimal transformations
(\ref{2.5K}), with $\xi$ defined by (\ref{xiAnizK}), into equation  (\ref{LPPref1}) with subsequent linearizing with respect to $a$ and using equation (\ref{LPPref}) to eliminate $dT^2$ yields
\begin{eqnarray}\label{DEAnisK}
\fl\left(-K c^2\tau_X
+\left(1-K^2\right)\left(K+c\tau_T\right)+\kappa\left(K\right) c
K\right)dX^2  \nonumber \\
+c \left(c^2\tau_X+c K \tau_T+1+K^2-\kappa\left(K\right) c\right)dX
dT  \nonumber \\
+\left(K+c\tau_T-c\eta_Y\right)dY^2
+\left(K+c\tau_T-c\zeta_Z\right)dZ^2
-c\left(\eta_Z+\zeta_Y\right)dY dZ=0
\end{eqnarray}
where subscripts denote differentiation with respect to the
corresponding variable. In view of arbitrariness of the
differentials $dX$, $dY$, $dZ$ and, $dT$, the equality (\ref{DEAnisK})
can be valid only if the coefficients of all the monomials in
(\ref{DEAnisK}) vanish which results in an overdetermined system of
determining equations for the group generators.

The generators $\tau$, $\eta$ and $\zeta$ found from the
determining equations yielded by (\ref{DEAnisK}) are
\begin{equation}\label{GenAnisK}
\eqalign{\tau=-\frac{1-K^2-\kappa\left(K\right) c}{c^2}X-\frac{2K}{c}T+c_2,\cr
\eta=-\frac{K}{c}Y+\omega Z+c_3,\quad\zeta=-\frac{K}{c}Z-\omega Y+c_4}
\end{equation}
where $c_2$, $c_3$ and $c_4$ are arbitrary constants. The
common kinematic restrictions that one event is the spacetime
origin of both frames and that the $x$ and $X$ axes slide along
another can be imposed to make the constants $c_2$, $c_3$ and
$c_4$ vanishing (space and time shifts are eliminated). In
addition, it is required that the $(x,z)$ and $(X,Z)$ planes
coincide at all times which results in $\omega=0$ and so excludes
rotations in the plane $(y,z)$.

The finite transformations are determined by solving the Lie equations which, after rescaling the group parameter as $\hat a=a/c$ together with
$\hat \kappa=\kappa c$ and omitting hats afterwards, take the forms
\begin{eqnarray}
\label{LEAnis0} &&\frac{d k(a)}{d
a}=\kappa\left(k\left(a\right)\right);\quad  k(0)=K,\\
\label{LEAnis1K}&&\frac{d x(a)}{d a}=-c t(a), \quad \frac{d \left(c t\left(a\right)\right)}{d
a}=-\left(1-k\left(a\right)^2-\kappa\left(k\left(a\right)\right)\right)x(a)-2 k(a) c t\left(a\right),\\
\label{LEAnis2P}&& \frac{d y(a)}{d a}=-k(a) y(a),\quad \frac{d
z(a)}{d a}=-k(a) z(a);\\
\label{ICK}&&x(0)=X,\; t(0)=T,\; y(0)=Y,\; z(0)=Z.
\end{eqnarray}
Because of the arbitrariness of $\kappa\left(k\left(a\right)\right)$, the solution of the system of equations
(\ref{LEAnis0}), (\ref{LEAnis1K}) and (\ref{LEAnis2P}) contains an arbitrary function $k(a)$. Using (\ref{LEAnis0})
to replace $\kappa\left(k\left(a\right)\right)$ in the second
equation of (\ref{LEAnis1K}) we obtain solutions of equations
(\ref{LEAnis1K}) subject to the initial conditions (\ref{ICK}) in the form
\begin{eqnarray}\label{txAnisP}
&& x=R\left(X \left(\cosh{a}+K\sinh{a}\right)-cT
\sinh{a}\right),\\
&& \label{txAnis1P} c t=R\Bigl(c T\left(
\cosh{a}-k\left(a\right)\sinh{a}\right)\nonumber\\
&&-X\left(\left(1-K k\left(a\right)\right)
\sinh{a}+\left(K-k\left(a\right)\right)\cosh{a}\right)\Bigr)
\end{eqnarray}
where $R$ is defined by
\begin{equation}\label{VarPhi}
R=\exp\left[-\int_0^a{k(\alpha)d\alpha}\right]
\end{equation}
The expression (\ref{VarPhi}) for the scale factor $R$ can be represented in a different form using equation (\ref{LEAnis0}), as follows
\begin{equation}\label{VarPhiaa}
R=\exp\left[-\int_K^k{\frac{p}{\kappa (p)}d p}\right]
\end{equation}

To complete the derivation of the transformations the group
parameter $a$ is to be related to the velocity $v$ using the
condition
\begin{equation}\label{2.8aK}
x=0 \quad \mathrm{for} \quad X=v T
\end{equation}
which yields
\begin{equation}\label{GrPAnisP}
a=\frac{1}{2}\ln{\frac{1+\beta - K\beta}{1-\beta-K\beta}};\qquad
\beta=\frac{v}{c}
\end{equation}
Substituting (\ref{GrPAnisP}) into (\ref{txAnisP}) and
(\ref{txAnis1P}) yields
\begin{eqnarray}\label{TrAnisP}
&&
x=\frac{R}{\sqrt{\left(1-K\beta\right)^2-\beta^2}}\left(X-c
T \beta\right),\nonumber \\
&&ct=\frac{R}{\sqrt{\left(1-K\beta\right)^2-\beta^2}}\left(c
T
\left(1-K\beta-k\beta\right)-X\left(\left(1-K^2\right)\beta+K-k\right)\right)
\end{eqnarray}
where $k$ is the value of $k(a)$ calculated for $a$ given by (\ref{GrPAnisP}).

 Solving equations
(\ref{LEAnis2P}) and using (\ref{GrPAnisP}) in the result yields
\begin{equation}\label{YZAnis2P}
y=R Y,\quad z=R Z
\end{equation}

Calculating the interval
\begin{equation}\label{IntAnisP}
ds^2=c^2 dt^2-2k c\; dtdx-(1-k^2)dx^2-dy^2-dz^2
\end{equation}
with (\ref{TrAnisP}) and (\ref{YZAnis2P}) yields
\begin{equation}\label{TrIntAnisP}
ds^2=R^2 dS^2,\quad dS^2=c^2 dT^2-2K c\;
dTdX-(1-K^2)dX^2-dY^2-dZ^2
\end{equation}
Thus,  the interval
invariance of the standard relativity is replaced by \textit{conformal invariance} with the
conformal factor dependent on the relative velocity of the frames
and the size of anisotropy in the frame $S$.

Nevertheless, there exists a combination which is invariant under the transformations and can be considered as a counterpart of the interval of the standard special relativity.
It is evident that the expression (\ref{VarPhiaa}) for the scale factor $R$ can be represented in the form
\begin{equation}\label{VarPhib}
R=\frac{\lambda (k)}{\lambda (K)}
\end{equation}
where
\begin{equation}\label{GR3.1.2}
\lambda (k)=\exp\left[-\int_0^k{\frac{p}{\kappa (p)}d p}\right]
\end{equation}
Then it follows from  equations (\ref{IntAnisP}) -- (\ref{VarPhib}) that the combination
\begin{equation}\label{GR3.1.1}
d\tilde s ^2=\frac{1}{\lambda (k)^2}\left(c^2 dt^2-2k c\; dtdx-(1-k^2)dx^2-dy^2-dz^2\right)
\end{equation}
is invariant under the transformations.

Furthermore, introducing the new variables
\begin{equation}\label{GR3.1.3}
\tilde t=\frac{1}{c \lambda (k)}\left(c t-k x\right),\quad \tilde x=\frac{1}{\lambda (k)}x,\quad \tilde y=\frac{1}{\lambda (k)}y,\quad \tilde z=\frac{1}{\lambda (k)}z
\end{equation}
converts the invariant combination (\ref{GR3.1.1}) into the Minkowski interval
\begin{equation}\label{GR3.1.4}
d\tilde s^2=c^2 d\tilde t^2-d\tilde x^2-d\tilde y^2-d\tilde z^2
\end{equation}
while the transformations defined by (\ref{txAnisP}), (\ref{txAnis1P}) and (\ref{YZAnis2P}) take the form of rotations in the $(\tilde x, \tilde t)$ space (Lorentz transformations)
\begin{equation}\label{2.8}
\tilde x=\tilde X \cosh{a}-c\tilde T \sinh{a},\quad c \tilde t=c \tilde T \cosh{a}-X \sinh{a};\qquad
\tilde y=\tilde Y,\quad \tilde z=\tilde Z
\end{equation}

The transformations defined by equations (\ref{TrAnisP}), (\ref{YZAnis2P}) and (\ref{GrPAnisP}) contain an indefinite function $k(a)$. The scale factor $R$
also depends on that function. In the next section, it is shown that incorporating the existence of a privileged frame into the analysis yields a formulation in which, instead of $k(a)$,  a function $k=F\left(\bar{\beta}\right)$, expressing dependence of the anisotropy size on the velocity  $\bar\beta$ of a frame with respect to the privileged frame, figures. A form of the latter function can be defined using some physical arguments which allows further specify the transformations.

\subsection{Special relativity with a privileged frame}

In derivation of the transformations in the previous section, nothing distinguishes a privileged frame, in which $k=0$,  from others and the transformations from/to that frame are members of a group of transformations that are equivalent to others. In this section, the transformations are specified using
an argument, that anisotropy of the one-way speed of light in an inertial  frame is due to its motion with respect to a privileged frame.
The argument leads to the conclusion that the anisotropy parameter $k$ in a frame moving with respect to a privileged frame with velocity $\bar\beta=\bar{v}/c$ should be given by some (universal) function $k=F\left(\bar{\beta}\right)$ of that velocity. It follows from equations (\ref{LEAnis0}) and (\ref{GrPAnisP}) which imply that  $k=k\left(a\left(\beta,K\right),K\right)$ so that for the transformation from the privileged frame to a frame $s$ we have $k_s=k\left(a\left(\bar\beta_s,0\right),0\right)$ or $k_s=F\left(\bar{\beta_s}\right)$.

Next, consider three inertial reference frames $\bar{S}$, $S$ and $S'$. As in the preceding analysis,
the standard configuration, with the $y$- and $z$-axes of the three
frames being parallel and the relative motion being along the
common $x$-axis (and along the direction of the anisotropy vector), is assumed. The space and time coordinates and the anisotropy parameters in the frames $\bar{S}$, $S$ and $S'$ are denoted respectively as $\{\bar{x},\bar{y},\bar{z},\bar{t},\bar{k}\}$, $\{X,Y,Z,T,K\}$ and $\{x,y,z,t,k\}$. The frame $S'$ moves relative to $S$ with velocity $v$ and velocities of the frames $S$ and $S'$ relative to the frame $\bar{S}$ are respectively $\bar{v_1}$ and $\bar{v_2}$. The relation between $\bar{v_2}$, $v$ and $\bar{v_1}$ can be obtained from the equation expressing a group property of the transformations, as follows
\begin{equation}\label{S1}
a_2=a_1+a
\end{equation}
where $a_2$, $a_1$ and $a$ are the values of the group parameter corresponding to the transformations from $\bar{S}$ to $S'$, from $\bar{S}$ to $S$  and from $S$ to $S'$ respectively. Those values are expressed through the velocities and the anisotropy parameter values by a properly specified equation (\ref{GrPAnisP}) which, upon substituting into equation (\ref{S1}), yields
\begin{equation}\label{S2}
\frac{1}{2}\ln{\frac{1+\bar{\beta_2} - \bar{k}\bar{\beta_2}}{1-\bar{\beta_2}-\bar{k}\bar{\beta_2}}}=\frac{1}{2}\ln{\frac{1+\bar{\beta_1} - \bar{k}\bar{\beta_1}}{1-\bar{\beta_1}-\bar{k}\bar{\beta_1}}}+\frac{1}{2}\ln{\frac{1+\beta - K\beta}{1-\beta-K\beta}}
\end{equation}
where
\begin{equation}\label{S3}
\bar{\beta_2}=\frac{\bar{v_2}}{c},\quad \bar{\beta_1}=\frac{\bar{v_1}}{c},\quad \beta=\frac{v}{c}
\end{equation}
Exponentiation of equation (\ref{S2}) yields
\begin{equation}\label{S4}
\bar{\beta_2}=\frac{\bar{\beta_1}+\beta\left(1-\left(\bar{k}+K\right)\bar{\beta_1}\right)}
{1+\beta\left(\bar{k}-K+\left(1-\bar{k}^2\right)\bar{\beta_1}\right)}
\end{equation}

Let us now choose the frame $\bar{S}$ to be a privileged frame.
Then, $\bar{k}=0$ and for the frames $S$ and $S'$ we have
\begin{equation}\label{S5}
K=F\left(\bar{\beta_1}\right),\quad k=F\left(\bar{\beta_2}\right)\quad \mathrm{or}\quad \bar{\beta_1}=f\left(K\right),\quad \bar{\beta_2}=f\left(k\right)
\end{equation}
where $\bar{\beta}=f\left(k\right)$ is a function inverse to $F\left(\bar{\beta}\right)$. Using (\ref{S5}) in (\ref{S4}) together with $\bar{k}=0$ yields
\begin{equation}\label{S6}
k=F\left(\frac{f\left(K\right)+\beta\left(1-K f\left(K\right)\right)}
{1+\beta\left(-K+f\left(K\right)\right)}\right)
\end{equation}
For a known function $F(\bar\beta)$ (and so for known $f\left(k\right)$), the relation (\ref{S6}) defines the anisotropy parameter $k$ in the frame $S'$ as a function of the anisotropy parameter $K$ in the frame $S$ and the relative velocity $\beta$ of the frames. and thus defines a form of the transformation of the anisotropy parameter. With $\beta$  expressed from (\ref{GrPAnisP}), as follows
\begin{equation}\label{S9}
\beta=\frac{\sinh{a}}{K\sinh{a}+\cosh{a}}
\end{equation}
equation (\ref{S6}) defines $k$ as a function of a group parameter and so allows to calculate the scale factor $R$ from (\ref{VarPhi}).

Alternatively, the relation (\ref{S6}) can be used for defining a form of the group generator $\kappa (k)$. Representing (\ref{S6}) in the form
\begin{equation}\label{S6a}
f\left(k\left(K;a\right)\right)=\frac{f\left(K\right)+\beta(a)\left(1-K f\left(K\right)\right)}
{1+\beta(a)\left(-K+f\left(K\right)\right)}
\end{equation}
substituting  (\ref{S9}) for $\beta(a)$ and differentiating the result with respect to $a$, with $\partial k(K;a)/\partial a$ separated,  yields
\begin{equation}\label{B2a}
\frac {\partial k(K;a)}{\partial a}=\frac {1-f(K)^2}{\left(\cosh{a}+f\left(K\right)\sinh{a}\right)^2
f'\left(k\left(a\right)\right)}
\end{equation}
Then the relation (\ref{S6a}), with $\beta$ substituted from (\ref{S9}), is used again to express $f(K)$ through  $f(k)$ and $a$. Substituting that expression into (\ref{B2a}) yields
\begin{equation}\label{B2}
\frac {d k(a)}{d a}=\frac {1-f^2\left(k\left(a\right)\right)}{f'\left(k\left(a\right)\right)}
\end{equation}
Equation (\ref{B2}) is the Lie equation defining (with the initial condition $k(0)=K$) the group transformation $k(K;a)$ which implies that the expression on the right-hand side is the group generator
\begin{equation}\label{B3}
\kappa (k)=\frac{1-f^2(k)}{f'(k)}
\end{equation}

A form of the function $F\left(\bar{\beta}\right)$ as an expansion in series of $\bar\beta$ can be defined based on the argument that the expansion should not contain even powers of $\bar\beta$ since it is expected that a direction of the anisotropy vector changes to the opposite if a direction of motion with respect to a privileged frame is reversed: $F\left(\bar{\beta}\right)=-F\left(-\bar{\beta}\right)$. In particular, with accuracy up to the third order in $\bar{\beta}$, the dependence of the anisotropy parameter on the velocity with respect to a privileged frame can be approximated by
\begin{equation}\label{S7}
k=F\left(\bar{\beta}\right)\approx b \bar{\beta}, \quad \bar{\beta}=f\left(k\right)\approx k/b
\end{equation}
Then using (\ref{S7}) in (\ref{S6}) yields
\begin{equation}\label{S8}
k=\frac{b\left(K+\beta\left(b-K^2\right)\right)}{b+\beta K\left(1-b\right)}
\end{equation}
which is the expression to be substituted for $k$ into (\ref{TrAnisP}).
To calculate the scale factor by (\ref{VarPhi}), $\beta(a)$ defined by (\ref{S9}) is substituted into
(\ref{S8}) to give
\begin{equation}\label{S10}
k(a)=\frac{b\left(K \cosh{a}+b \sinh{a}\right)}{K\sinh{a}+b\cosh{a}}
\end{equation}
Then using (\ref{S10}) in (\ref{VarPhi}), with (\ref{GrPAnisP}) substituted for $a$ in the result,  yields
\begin{equation}\label{S11}
R=\left(\frac{b^2\left(1+\beta\left(1-K\right)\right)\left(1-\beta\left(1+K\right)\right)}
{\left(b+\beta K\left(1-b\right)\right)^2}\right)^{\frac{b}{2}}
\end{equation}

Thus, after the specification, the transformations between inertial frames incorporating anisotropy of light propagation are defined by equations (\ref{TrAnisP}) and (\ref{YZAnis2P}) with $k$ given by (\ref{S8}) and the scale factor given by (\ref{S11}). It is readily checked that the specified transformations satisfy the correspondence principle. All the equations contain only one undefined parameter, a universal constant $b$.

It should be clarified that, although the specification relies on the approximate relation (\ref{S7}), the transformations themselves, even with $k$ and $R$ defined by (\ref{S8}) and (\ref{S11}), are \textit{not} approximate and they do possess the group property.  The transformations (\ref{TrAnisP}) and (\ref{YZAnis2P}) form a group, even with $k(a)$ (or $k(K,\beta)$) undefined, provided that the transformation of $k$ obeys the group property. Since the relation (\ref{S6}), defining that transformation, is  a particular case of the relation (\ref{S4}) obtained from equation (\ref{S1}) expressing the group property, the transformation of $k$ satisfies the group property with any form of the function $k=F(\bar \beta)$, and, in particular, with that defined by (\ref{S7}). It can be demonstrated by a straightforward check or, alternatively, we can calculate the group generator $\kappa(k)$ from (\ref{B3}) using the expression  (\ref{S7}) for $f(k)$ which yields
\begin{equation}\label{S14} \kappa(k)=b-\frac{k^2}{b}
\end{equation}
Then solving the initial value problem
\begin{equation}\label{S15}
\frac{d k(a)}{d
a}=b-\frac{k(a)^2}{b},\quad k(0)=K
\end{equation}
yields (\ref{S10}), as expected, while using (\ref{S14}) in (\ref{VarPhiaa}), with (\ref{S8}) substituted for $k$ in the result, yields (\ref{S11}).

With the expression (\ref{S14}) for $\kappa(k)$, based on the approximation (\ref{S7}), the factor $\lambda (k)$  is calculated from (\ref{GR3.1.2}) as
\begin{equation}\label{GR3.1.5}
\lambda (k)=\left(1-\frac{k^2}{b^2}\right)^{b/2}
\end{equation}
If equation   (\ref{S7}) is introduced into (\ref{GR3.1.5}) the factor $\lambda (k)$ becomes a function $B (\bar\beta)$ of the frame velocity $\bar \beta$ relative to a privileged frame
\begin{equation}\label{GR3.1.6}
B (\bar\beta)=\left(1-\bar\beta^2\right)^{b/2}
\end{equation}
With the same order of approximation as that in (\ref{S7}), the expression (\ref{GR3.1.6}) for $B (\bar\beta)$  can be represented as
\begin{equation}\label{GR3.1.7}
B (\bar\beta)=1-\frac{b}{2}\bar\beta^2
\end{equation}

An expression for the factor $B(\bar\beta)$ for arbitrary $F(\bar\beta)$ is derived in Appendix A.

\section{General relativity}


The basic principle of general relativity (The Equivalence Principle) asserts that at each point of spacetime it is possible to choose a 'locally inertial' coordinate system in which the effects of gravitation are absent and the special theory of relativity is valid.
It is evident that the principle, that it is always possible to choose a locally inertial frame in which objects obey Newton's first law, is valid independently of the law of propagation of light assumed. It implies that
the equivalence principle can be applied when the processes in the locally inertial frames are governed by the modified special relativity based on invariance of anisotropic equation of light propagation.
Developing the general relativity using the equivalence principle in the latter case seems problematic since the interval is not invariant but conformally modified under the transformations. Nevertheless, the complete apparatus of general relativity can be applied based on that there exists the invariant combination  (\ref{GR3.1.1}) which takes the form of the Minkowski interval upon the change of variables (\ref{GR3.1.3}).
Thus, the general relativity equations in arbitrary coordinates $(x^0,x^1,x^2,x^3)$ are valid if
the  locally inertial coordinates $(\xi^0,\xi^1,\xi^2,\xi^3)$ are defined as
\begin{equation}\label{GR3.1.8}
\xi^0=c \tilde t,\; \xi^1=\tilde x,\; \xi^2=\tilde y,\; \xi^3= \tilde z
\end{equation}
where $\tilde t$, $\tilde x$, $\tilde y$ and $\tilde z$ are defined in (\ref{GR3.1.3}),
and the invariant spacetime distance squared
$ds^2=g_{ik}dx^idx^k$ is equal to $d\tilde s^2=\eta_{ik}d\xi^id\xi^k$ (repeated indices are summed and the common notation $\eta_{ik}$ is used for the  Minkowski metric). However, in the calculation of physical effects, the 'true' time and space intervals in the 'physical' variables $(t, x, y, z)$
are to be used.

It is worthwhile to remark that it does not influence a validity of the arguments based on the small velocity limits that are commonly used in developing
a framework of the general relativity. In those limiting arguments, only the first order in $\beta=v/c$ terms are considered while the difference between the 'locally inertial' coordinates  $(\tilde t, \tilde x, \tilde y, \tilde z)$ and 'physical' coordinates $(t, x, y, z)$ is, according to (\ref{GR3.1.3}) and (\ref{GR3.1.7}), of the second order in $\beta$. Note, in addition, that the equation of a freely moving particle in a locally inertial frame, which plays an important role in developing a paradigm of general relativity, takes, in the 'locally inertial' coordinates $\xi^i$ defined by (\ref{GR3.1.8}) and (\ref{GR3.1.3}),  the same form as in the physical coordinates $(t, x, y, z)$, as follows
\begin{equation}\label{GR3.1.9}
\frac{d^2\xi^i}{d\tau^2}=0; \quad d\tau^2=\frac{1}{c^2}\eta_{ik}d\xi^id\xi^k
\end{equation}
where $\tau$ is the proper time.


In what follows, the notation for 'physical' coordinates $(t, x, y, z)$ is changed to $(t^{\ast}, x^{\ast}, y^{\ast}, z^{\ast})$ to leave freedom for using $(c t, x, y, z)$ instead of $(x^0,x^1,x^2,x^3)$ in the contexts where it is traditionally done in the literature. For the sake of convenience, equations (\ref{GR3.1.3}) relating the physical coordinates $(t^{\ast}, x^{\ast}, y^{\ast}, z^{\ast})$ to the 'locally inertial' coordinates $(\xi^0,\xi^1,\xi^2,\xi^3)$  are rewritten below with taking into account the relations (\ref{GR3.1.8}) and (\ref{GR3.1.3}), as follows
\begin{equation}\label{GR3.1.10}
t^{\ast}=\frac{1}{c}B (\bar\beta)\left(\xi^0+k \xi^1\right),\; x^{\ast}=B (\bar\beta)\xi^1,\; y^{\ast}=B (\bar\beta)\xi^2,\; z^{\ast}=B (\bar\beta)\xi^3
\end{equation}
where $\bar \beta$ is the velocity of a locally inertial (freely falling) observer relative to a privileged frame and, with an accuracy up to terms of order $\beta^2$, the factor $B(\beta)$ can be approximated by (\ref{GR3.1.7}).

Let us now determine the relations of the 'true' time and space intervals to the coordinates  $(x^0,x^1,x^2,x^3)$. First, recall the relation of the proper time interval $d\tau$ to the interval  $dx^0$. That relation is obtained by considering two infinitesimally separated events occurring at one and the same point in space $dx^1=dx^2=dx^3=0$ (see,  e.g., \cite{LL}) which yields
\begin{equation}\label{GR3.1.11}
d\tau=\frac{1}{c}\sqrt{g_{00}}dx^0
\end{equation}
Next,  using the relation $d\tau=d\xi^0/c$ following from invariance of the spacetime distance, together with (\ref{GR3.1.10}) (with $d\xi^1=0$) and (\ref{GR3.1.11}), in calculation of the 'true' proper time interval $dt^{\ast}$ yields
\begin{equation}\label{GR3.1.12}
dt^{\ast}=B (\bar\beta)d\tau=\frac{1}{c}B (\bar\beta)\sqrt{g_{00}}dx^0
\end{equation}
To obtain an expression for the element $dl^{\ast}$ of 'true' spatial distance
consider, following \cite{LL}, a light signal sent from some point B in space with coordinates $x^{\alpha}+d x^{\alpha}$ to a point A with coordinates $x^{\alpha}$ (here and below Greek indices run from 1 to 3, while Latin indices run from 0 to 3)  and then back over the same part. The time  required for this (as measured at the point B), when multiplied by $c$, is twice the distance between the two points. Determining the interval $d x^0$ between the departure of the signal and its return to B (see \cite{LL}) yields
\begin{equation}\label{GR3.1.13}
dx^0=\frac{2}{g_{00}}\sqrt{\left(g_{0\alpha}g_{0\beta}-g_{\alpha \beta}g_{00}\right)dx^{\alpha}dx^{\beta}}
\end{equation}
The corresponding interval of the 'true' proper time $dt^{\ast}$ is obtained using (\ref{GR3.1.12}) and the distance $dl^{\ast}$ between the two points is obtained by multiplying it by $c/2$ which yields
\begin{equation}\label{GR3.1.14}
dl^{\ast}=B (\bar\beta)\sqrt{\gamma_{\alpha\beta}dx^{\alpha}dx^{\beta}},\quad \gamma_{\alpha\beta}=-g_{\alpha \beta}+\frac{g_{0\alpha}g_{0\beta}}{g_{00}}
\end{equation}
In view of the fact that the time and the distance intervals are modified by the same factor $B (\bar\beta)$, the expression for the proper velocity of a particle $v=dl^{\ast}/dt^{\ast}$
does not include that factor and so the proper velocity is calculated in a usual way.


Below we apply the modified GR to cosmology leaving aside other possible applications. The problem of a gravitational collapse of a dustlike sphere, which, in some aspects, is related to cosmological issues, is considered in Appendix B.

\section{Cosmological models}

\subsection{General framework}

Modern cosmological models are based on the assumption that the universe appear isotropic to "typical" freely falling observers, those that move with the average velocity of typical
galaxies in their respective neighborhoods.
It is also assumed that such the typical (privileged) Lorentzian frame in which the universe appears isotropic coincides more or less with our own galaxy.

The metric derived on the basis of isotropy
and homogeneity (the \textit{Robertson4--Walker metric}) has the form
\begin{equation}\label{GR3.1.27}
ds^2=dt^2-a^2(t)\left(\frac{dr^2}{1-K_c r^2}+r^2 d\Omega\right),\quad d\Omega=d\theta^2+\sin^2\theta d\phi^2
\end{equation}
where a co-moving reference system, moving at each point of space along with the matter located at that point, is used. This implies that the coordinates $(r,\theta, \phi)$ are unchanged for each typical observer (presumably located at a galaxy).
In (\ref{GR3.1.27}), and in what follows, the system of units in which the speed of light is equal to unity, is used. The time
coordinate $x^0=t$ is the synchronous proper time at each point of space.
The constant $K_c$ (this notation is used, instead of common $k$ or $K$, to avoid confusion with the symbols for the anisotropy parameter)
 by a suitable choice of units for $r$ can be chosen to have the value $+1$, $0$, or $-1$. Introducing, instead of $r$, the radial coordinate $\chi$ by the relation $r=S(\chi)$ with
\begin{equation}
\label{GR3.1.27a}
S(\chi)=\cases{\sin \chi&for $K_c=1$\\
\sinh \chi&for $K_c=-1$\\
\chi&for $K_c=0$\\}
\end{equation}
converts (\ref{GR3.1.27})
into the form
\begin{equation}\label{GR3.1.28}
ds^2=dt^2-a^2(t)\left[d\chi^2+S^2(\chi)\; d\Omega\right]
\end{equation}
Next,
let us introduce, in place of the time $t$, the \textit{conformal
time} $\eta$ defined by
\begin{equation}\label{GR3.1.29}
dt=a(t)d\eta
\end{equation}
Then the function $a(t)$ may be treated as a function of $\eta$ (for what follows, it is convenient to leave the same notation for that function) and $ds^2$ can be written as
\begin{equation}\label{GR3.1.30}
ds^2=a^2(\eta)\left[d\eta ^2-d\chi^2-S^2(\chi)\; d\Omega\right]
\end{equation}


\subsection{The red shift}

The information about the scale factor $a(t)$ in the Robertson–Walker metric can be obtained from observations of shifts in frequency of light emitted by distant sources. To calculate such frequency shifts let us consider the propagation of a light ray in an isotropic space with the metric (\ref{GR3.1.30}) adopting a coordinate system
in which we are at the center of coordinates $\chi=0$ and the source is at the point with a coordinate $\chi=\chi_1$. A light ray propagating along the radial direction obeys the equation $d\eta ^2-d\chi^2=0$. For a light ray coming toward the origin from the source, that equation gives
\begin{equation}\label{GR3.1.34}
\chi_1=-\eta_1+\eta_0
\end{equation}
where $\eta_1$ corresponds to the moment of emission $t_1$ and $\eta_0$ corresponds to the moment of observation $t_0$.
Let $dt_1$ is the time interval between departure of subsequent light signals from the point $\chi=\chi_1$ and $dt_0$ is the time interval between arrivals of these light signals to the observer at the point $\chi=0$. It follows from equation (\ref{GR3.1.34}) that the corresponding increments of the variable $\eta$ are equal to each other $d\eta_1=d\eta_0$ (the co-moving coordinate $\chi$ is time-independent) which, upon using equation
(\ref{GR3.1.29}), gives
\begin{equation}\label{GR3.1.35}
\frac{dt_1}{a(\eta_1)}=\frac{dt_0}{a(\eta_0)}
\end{equation}
Next, the time intervals $dt_1$ and $dt_0$ are to be related to the intervals of physical time $dt_1^{\ast}$ and $dt^{\ast}_0$ using the relation
\begin{equation}\label{GR3.1.35a}
dt^{\ast}=B(\bar\beta) dt,\quad B(\bar\beta)=1-\frac{b}{2}\bar\beta^2
\end{equation}
which requires calculation of the velocity $\bar \beta$ with respect to a privileged frame. It is evident that for a 'typical' freely falling observer
all other typical observers (galaxies) are moving in radial direction and the privileged frame for such an observer is one in which the distribution of the velocities of galaxies appears isotropic. Thus, the observer at the origin of coordinates is at rest and  the source is moving with the velocity $\bar\beta=\bar\beta_1(\chi_1,\eta_1)$ with respect to the privileged frame so that, with the use of (\ref{GR3.1.12}), equation  (\ref{GR3.1.35}) takes the form
\begin{equation}\label{GR3.1.36}
\frac{dt_1^{\ast}}{a\left(\eta_1\right)B \left(\bar\beta_1\right)}=\frac{dt_0}{a(\eta_0)},\quad B(\bar\beta_1)=1-\frac{b}{2}\bar\beta_1^2
\end{equation}
If the "signals" are subsequent wave crests, the observed frequency $\nu_0=1/dt_0$ is related to the frequency of the emitted light $\nu_1=1/dt_1^{\ast}$ by
\begin{equation}\label{GR3.1.37}
\frac{\nu_0}{\nu_1}=\frac{a\left(\eta_1\right)B \left(\bar\beta_1\right)}{a(\eta_0)}
\end{equation}
where the frequency $\nu_1$ of a spectral line coincides with that observed in terrestrial laboratories. The \textit{red-shift parameter} $z$ is defined by
\begin{equation}\label{GR3.1.380}
z
=\frac{\nu_1}{\nu_0}-1
\end{equation}
With the use of equations (\ref{GR3.1.37}) and (\ref{GR3.1.34}), the relation (\ref{GR3.1.380}) takes the form
\begin{equation}\label{GR3.1.38}
z=\frac{a(\eta_0)}{a(\eta_0-\chi_1)B(\bar\beta_1)}-1
\end{equation}

\subsection{The red-shift versus luminosity distance relation}

The relation expressing the \textit{Luminosity Distance} of a cosmological source in terms of its redshift $z$ is one of the fundamental relations in cosmology. It has been exploited to get information about the time evolution of the expansion rate.
Let $dE_1$ is the energy emitted by the source during the interval of ('physical') time $dt_1^{\ast}$. Then the absolute luminosity $L$ of the source is defined by
\begin{equation}\label{GR3.1.38a}
L=\frac{dE_1}{dt_1^{\ast}}=\frac{dE_1}{B(\bar\beta_1) dt_1}
\end{equation}
where the relation (\ref{GR3.1.35a}) has been used. Let us assume that the energy
is emitted isotropically and imagine the luminous
object to be surrounded with a sphere whose radius is equal to the distance
between the source and the observer. With the metric (\ref{GR3.1.27}), the area of the surface of the sphere
at the moment of observation $t_0$ is equal to $4\pi a^2(t_0) S^2(\chi_1)$. Then the apparent luminosity $l$ (the energy passed per unit time per unit area of the surface of the sphere) is calculated as
\begin{equation}\label{GR3.1.39}
l=\frac{dE_0/dt_0}{4\pi a^2(t_0) S^2(\chi_1)}
\end{equation}
where $dE_0$ is the total energy passed through the surface during the time interval $dt_0$. The time interval $dt_0$ is related to the time interval $dt_1$, within which the energy was emitted, by equation (\ref{GR3.1.35}) and the energy $dE_0$ received by the surface of the sphere is related to the emitted energy $dE_1$ by
\begin{equation}\label{GR3.1.39a}
\frac{dE_0}{dE_1}=\frac{h\nu_0 dN}{h\nu_1 dN}=\frac{\nu_0}{\nu_1}
\end{equation}
where $h\nu_0$ and $h\nu_1$ are the energies of the individual photons received by the observer and emitted by the source and $dN$ is the number of photons emitted during the time interval $dt_1$. The luminosity distance $d_L$ is defined based on the relation from euclidian geometry $l=L/(4\pi d_L^2$) by  \cite{LL},\cite{SW1},\cite{SW2}
\begin{equation}\label{GR3.1.39b}
d_L=\sqrt{\frac {L}{4\pi l}}
\end{equation}
Substituting (\ref{GR3.1.38a}) for $L$ and (\ref{GR3.1.39}) for $l$ with a subsequent use of equations (\ref{GR3.1.35}), (\ref{GR3.1.39a}) and (\ref{GR3.1.37}) yields
\begin{equation}\label{GR3.1.39c}
d_L=\frac{a^2(t_0)S(\chi_1)}{a(t_1)B(\bar\beta_1)}=\frac{a^2(\eta_0)S(\chi_1)}{a(\eta_0-\chi_1)B(\bar\beta_1)}
\end{equation}
By eliminating $a(\eta_0-\chi_1)$ using (\ref{GR3.1.38}), as it is usually done, another form of the relation for $d_L$ is obtained, namely
\begin{equation}\label{GR3.1.40}
d_L=a(\eta_0)(1+z)S(\chi_1)
\end{equation}
This relation  coincides with a common form of the relation for $d_L$  \cite{LL},\cite{SW1},\cite{SW2}. Nevertheless, even though it does not contain the factor $B(\bar\beta_1)$, the dependence of $d_L$ on $z$ obtained by eliminating $\chi_1$ from equations (\ref{GR3.1.40}) and (\ref{GR3.1.38}) will differ from the common one since the relation (\ref{GR3.1.38}) for $z$ does contain the factor $B(\bar\beta_1)$.

To calculate the factor $B(\bar\beta_1)$
the value of $\bar\beta_1$ is to be determined.
 The proper radial velocity $\bar\beta$ of a remote object cannot be determined using the expression (\ref{GR3.1.14}) for the distance passed by the object since the comoving coordinates $x^{\alpha}=\{\chi,\theta,\phi\}$ do not change during the particle motion while (\ref{GR3.1.14}) deals with the coordinate increments $dx^{\alpha}$.
 The particle velocity with respect to the center can be calculated as $\bar \beta=\partial L(\chi,t)/\partial t$ where $L(\chi,t)$ is the proper distance of the particle with the radial coordinate $\chi$ to the center. Commonly the quantity $d(\chi,t)=a(t)\chi$ is called the proper distance \cite{SW1},\cite{SW2} but, in fact, it is not the proper distance to the center as that relation is obtained by integrating the 'radial' line element $dl=a(t)d\chi$ for constant $t$ which corresponds to simultaneous observation of all the points along the path of integration and so is physically not feasible \cite{LL},\cite{SW1}. Nevertheless, the relation $L(\chi,t)=a(t)\chi$ provides a small $\chi$ approximation for the proper distance
\cite{LL}. The corresponding relation for the velocity of an object with respect to the center (with respect to a privileged frame) is
\begin{equation}\label{GR3.1.41}
\bar \beta=\frac{da(t)}{dt}\Bigg\vert_{t=t_0}\chi=a'(t_0)\chi=\frac{a'(\eta_0)}{a(\eta_0)}\chi
\end{equation}
Here, and in what follows, the derivatives with respect to $t$ are converted into derivatives with respect to $\eta$ using equation (\ref{GR3.1.29}).
Although the relation (\ref{GR3.1.41}) contains only a term of the first order in $\chi$, the approximation is sufficiently accurate. Since the expression (\ref{GR3.1.36}) for the factor $B(\bar\beta_1)$ depends on $\bar\beta_1^2$, introducing  the approximation (\ref{GR3.1.41}) into  (\ref{GR3.1.36}) makes it valid up to the terms of the order $\chi_1^2$.
Within the same accuracy, the factor $1/B (\bar\beta_1)$ can be expressed as $1/B (\bar\beta_1)=1+(b/2)\bar\beta_1 ^2$. Then incorporating this relation and the relation (\ref{GR3.1.41}) in (\ref{GR3.1.38}) yields
\begin{equation}\label{GR3.1.42}
z=\frac{a(\eta_0)}{a(\eta_0-\chi_1)}\left(1+b\frac{a'^2(\eta_0)}{2a^2(\eta_0)}\chi_1^2\right)-1
\end{equation}

To derive the relation between luminosity distance and red-shift as a power series, $z$ and $d_L$ defined by equations  (\ref{GR3.1.42}) and (\ref{GR3.1.40}) are expanded in series of $\chi_1$ (in the literature, the series in the 'look-back time' $t_0-t_1$ are commonly used in that derivation), which, upon retaining terms up to the order $\chi_1^2$, yields
\begin{equation}\label{GR3.1.43}
z=H_0 a(\eta_0)\chi_1+\frac{1}{2}H_0^2a^2(\eta_0)(1+b+q_0)\chi_1^2+\cdots
\end{equation}
\begin{equation}\label{GR3.1.44}
d_L=a(\eta_0)(1+z)\chi_1+\cdots
\end{equation}
where $H_0$ and $q_0$ are defined by
\begin{equation}
\label{GR3.1.45}
H_0=\frac{a'(t_0)}{a(t_0)}=\frac{a'(\eta_0)}{a^2(\eta_0)},\quad q_0=-\frac{1}{H_0^2 a(t_0)}\frac{d^2a(t)}{dt^2}\Bigg\vert_{t=t_0}=1-\frac{a''(\eta_0)}{H_0^2 a^3(\eta_0)}
\end{equation}
Equation (\ref{GR3.1.43}) can be inverted to give the source coordinate $\chi_1$ as a power series in the redshift
\begin{equation}\label{GR3.1.46}
\chi_1=\frac{1}{H_0 a(\eta_0)}\left(z-\frac{1}{2}\left(1+b+q_0\right)z^2+\cdots\right)
\end{equation}
Substituting (\ref{GR3.1.46}) into (\ref{GR3.1.44}) gives the luminosity distance as a power series
\begin{equation}\label{GR3.1.47}
d_L=H_0^{-1}\left(z+\frac{1}{2}\left(1-q_0-b\right)z^2+\cdots\right)
\end{equation}

The relation (\ref{GR3.1.47}) is quite general in a sense that it has been derived using the Robertson-Walker metric based solely on the assumptions of isotropy and homogeneity.  However, the expansion rate parameters $H_0$ and $q_0$ remain unspecified. To go further
one needs to consider the dynamics of the cosmological expansion by applying the gravitational field equations of Einstein. It allows to relate the parameters of the expansion
to the values of the cosmic energy density and
pressure and, upon making some tentative assumptions about constituents of the universe and their properties, to obtain theoretical predictions for the values of the parameters.

For a  matter-dominated cosmological model of the universe (Friedman model) based on the standard GR solving the gravitational field equations   yields the luminosity distance -- redshift relation of the form
\begin{equation}\label{GR3.1.48}
d_L=H_0^{-1}\left(z+\frac{1}{2}\left(1-q_0^{(D)}\right)z^2+\cdots\right)
\end{equation}
where the deceleration parameter $q_0^{(D)}$ is positive for all three possible values of the parameter $K_c$
which means that, in that model, the expansion of the universe is
decelerating. However, recent observations of Type Ia supernovae (SNIa), fitted into
the luminosity distance versus redshift relation of the form (\ref{GR3.1.48}), correspond to the deceleration parameter $q_0^{(D)}<0$ which indicates that the expansion of the universe is
accelerating. This result is interpreted as that the time evolution of the expansion rate cannot be described by a matter-dominated Friedmann-Robertson-Walker cosmological model of the universe. In order to explain the discrepancy within the context of general relativity, the dark energy,  a new component of the energy density with strongly negative pressure that makes the universe accelerate, is introduced  (see, e.g., \cite{SW2}).


The framework of the relativity with a privileged frame developed in the present study, which leads to the luminosity distance -- redshift relation of the form (\ref{GR3.1.47}), allows another interpretation of the results of observations with supernovae. According to (\ref{GR3.1.47}), the observed deceleration parameter  $q_0^{(D)}$ in the relation (\ref{GR3.1.48}) is $q_0^{(D)}=q_0+b$ where $q_0$ is the deceleration parameter of the Friedman-Robertson-Walker model. Since the parameter $b$ is expected to be negative, the observed negative values of $q_0^{(D)}$ do not exclude the Friedman dynamics with $q_0>0$ corresponding to the decelerating universe. Thus, within the framework developed in the present study, the acceleration problem can be naturally resolved.
With the value $q_0^{(D)}\approx -0.6$ found in observations, the parameter $b$ is estimated to be $b \approx -1.1$.

To make the analysis consistent, the dependence $d_L(z)$ is to be specified 
by relating the parameters to the values of the cosmic energy density and
pressure using the gravitational field equations while introducing alterations, that originate from relativity with a privileged frame, into the results. The fundamental Friedmann equation, which is obtained as a consequence of the Einstein field equations, can be written in the form (see, e.g., \cite{SW2})
\begin{equation}\label{GR3.1.49}
a'^2(t)=-K_c+\frac{8\pi G}{3}a^2(t)\rho(t)
\end{equation}
where $G$ is Newton's gravitational constant.
Applying this equation requires making
assumptions about the cosmic energy density $\rho$ and the form of "equation of state" giving the pressure $p$  as a function of the energy density.
The energy density $\rho(t)$ is usually assumed
to be a mixture of non-relativistic matter with equation of state $p=0$ and dark energy  with equation of state $p=w \rho$ while ignoring the relativistic matter (radiation). In the commonly accepted  $\Lambda$CDM model, the dark energy obeys equation of state with $w=-1$ (vacuum energy) which is equivalent to introducing into
Einstein's equation a cosmological constant $\Lambda$.
Then $\rho(t)$ is expressed as
\begin{equation}\label{GR3.1.50}
\rho (t)=\frac {3H_0^2}{8\pi G}\left(\Omega_{\Lambda}+\Omega_M x(t)^{-3}\right)
\end{equation}
where
\begin{equation}\label{GR3.1.51}
x(t)=\frac{a(t)}{a_0},\quad a_0=a(t_0)
\end{equation}
The parameters $\Omega_{\Lambda}$, $\Omega_M$ are defined by
\begin{equation}\label{GR3.1.52}
\Omega_{\Lambda}=\frac {\rho_{V0}}{\rho_c}, \quad \Omega_M=\frac {\rho_{M0}}{\rho_c};\qquad \rho_c=\frac{3H_0^2}{8\pi G}
\end{equation}
where $\rho_{V0}$ and $\rho_{M0}$ are the present energy densities in vacuum and non-relativistic matter and $\rho_c$
is the critical energy density. Using equations (\ref{GR3.1.50}) and (\ref{GR3.1.51}) in equation (\ref{GR3.1.49}) yields
\begin{equation}\label{GR3.1.53}
\left(x'\right)^2=H_0^2 x^2\left(\Omega_{\Lambda}+\Omega_M x^{-3}+\Omega_K x^{-2}\right)
\end{equation}
where
\begin{equation}\label{GR3.1.54}
\Omega_K=-\frac{K_c}{a_0^2 H_0^2}
\end{equation}
and the argument of $x$ is omitted for convenience of using the right-hand side of (\ref{GR3.1.53}) in an integral with respect to $x$, in what follows. Being evaluated at $t=t_0$ equation (\ref{GR3.1.53}) becomes
\begin{equation}\label{GR3.1.55}
\Omega_{\Lambda}+\Omega_M+\Omega_K=1
\end{equation}
The Friedmann equation (\ref{GR3.1.53}) allows us to calculate the radial coordinate $\chi_1$ of an object of a given redshift $z$. Equation (\ref{GR3.1.34}) defining  $\chi_1$  can be represented in the form
\begin{equation}\label{GR3.1.55a}
\chi_1=\eta_0-\eta_1=\int_{\eta_1}^{\eta_0}{d\eta}=\int_{t_1}^{t_0}{\frac{dt}{a(t)}}=\frac{1}{a_0}\int_{x_1}^{1}{\frac{dx}{x'x}}
\end{equation}
where $x'$ is a function of $x$ defined by the Friedmann equation  (\ref{GR3.1.53}) and $x_1=a(t_1)/a_0$. Then  using equation (\ref{GR3.1.53}) in (\ref{GR3.1.55a}) yields
\begin{equation}\label{GR3.1.55c}
\chi_1=\int_{x_1}^{1}{\frac{dx}{a_0H_0 x^2 \sqrt{\Omega_{\Lambda}+\Omega_M x^{-3}+\Omega_K x^{-2}}}}
\end{equation}

In the standard cosmology, equation (\ref{GR3.1.38})  (with $B(\bar\beta_1)=1$) provides a simple relation
\begin{equation}\label{GR3.1.56}
x_1=\frac{1}{1+z}
\end{equation}
so that (\ref{GR3.1.55c}) becomes a closed-form relation for $\chi_1(z)$. 
For a 'concordance' model, which is the flat space  $\Lambda$CDM model,  $\Omega_K=0$ and $\Omega_{\Lambda}=1-\Omega_M$ and then calculating the integral in (\ref{GR3.1.55c}), with $x_1$ defined by (\ref{GR3.1.56}), yields an expression for $\chi_{1}(z)$ in terms of a hypergeometric function $_2 F^1(a_1,a_2;a_3;x)$, as follows
\begin{eqnarray}\label{GR3.1.69}
\fl\chi_{1m}^{c}(z)=\frac{1}{\sqrt{1-\Omega_M^{\;c}}}\Biggl({_2F^1}\left(\frac{1}{3},\frac{1}{2};\frac{4}{3};\frac{\Omega_M^{\;c}}{\Omega_M^{\; c}-1}\right)\nonumber\\
+\left(1+z\right)
{_2F^1}\left(\frac{1}{3},\frac{1}{2};\frac{4}{3};\frac{\Omega_M^{\;c}\left(1+z\right)^3}{\Omega_M^{\;c}-1}\right)\Biggr) \quad \mathrm{where} \quad \chi_{1m}^{c}=a_0 H_0 \chi_1^{c}
\end{eqnarray}
Here and in what follows, quantities with a superscript $"c\;"$ refer to the concordance model, with the original notation secured for the corresponding quantities of the present model.
Then the luminosity distance is calculated as
\begin{equation}\label{GR3.1.70}
d_L^{\;c}(z)=\frac{1}{H_0}\left(1+z\right)\chi_{1m}^{c}(z)
\end{equation}
with $\chi_{1m}^{c}(z)$ given by (\ref{GR3.1.69}).

In the framework of the present analysis, expressing $\chi_1$ as a function of $z$ by combining equations (\ref{GR3.1.55c}) and (\ref{GR3.1.38}) becomes more complicated in view of the fact that $\bar\beta_1$, and so  the factor $B\left(\bar\beta_1\right)$,
depend on $\chi_1$.
Therefore determining $\chi_1$ for a given $z$ requires solving a system of two equations for $x_1$ and $\chi_1$, one of which is (\ref{GR3.1.55c}) and the second is
\begin{equation}\label{GR3.1.57}
z+1=\frac{1}{x_1 B\left(\bar\beta_1\left(\chi_1\right)\right)}
\end{equation}
with the properly specified function $B\left(\bar\beta_1\left(\chi_1\right)\right)$. If it is possible to invert the relation (\ref{GR3.1.55c}) to get an expression for  $x_1(\chi_1)$ then the problem reduces to a transcendental equation for $\chi_1(z)$ obtained by substituting $x_1(\chi_1)$  into (\ref{GR3.1.57}).

With the presumption, that in the cosmology based on the relativity with a privileged frame there is no need in introducing dark energy ($\Omega_{\Lambda}=0$), the relation
$\chi_1(x_1)$ can be obtained from (\ref{GR3.1.55c}) in an analytical form. Then inverting the result yields
\begin{equation}\label{GR3.1.58c}
x_1=\frac{1}{\Omega_K}\left[\sinh{\left(\frac{1}{2}\sqrt{\Omega_K}\chi_{1m}\right)}
-\sqrt{\Omega_K}\cosh{\left(\frac{1}{2}\sqrt{\Omega_K}\chi_{1m}\right)}\right]^2,\quad \chi_{1m}=a_0 H_0 \chi_1
\end{equation}
Although the form of this equation implies that $\Omega_K>0$, it is also  applicable to the cases of  $\Omega_K<0$ and $\Omega_K=0$. For $\Omega_K<0$,
the argument of the hyperbolic sine and cosine is imaginary, and using the relations $\sinh ix =i \sin x$ and $\cosh ix=\cos x$ yields a proper expression for $x_1$. Also, equation  (\ref{GR3.1.58c}) has a smooth limit for
$\Omega_K\rightarrow 0$, which gives the result for zero curvature.
Substituting (\ref{GR3.1.58c}) into (\ref{GR3.1.57}) yields a transcendental equation for $\chi_1$ if the expression for the factor $B\left(\bar\beta_1\left(\chi_1\right)\right)$ is specified. In what follows, a solution of the equation for $\chi_1$ is represented as series in $z$. To provide a sufficient accuracy for a reliable comparison of the results with observational data in this and the next subsections, the third order in $z$ terms are to be taken into account. Since $\chi_1$ is of the order of $z$ it requires also including the third order in $\chi_1$ terms into the expression for the factor $B\left(\bar\beta_1\left(\chi_1\right)\right)$. In that context, the approximation
\begin{equation}\label{GR3.1.60b}
B(\bar\beta_1)=1-\frac{b}{2}\bar\beta_1^2
\end{equation}
used in the second order calculations of the previous section is sufficient since, according to (\ref{B7}), the next order term in $B(\bar\beta_1)$ is of the order of $\bar\beta_1^4$ while $\bar\beta_1$ is of the order of $\chi_1$. However, equation (\ref{GR3.1.41}) defining $\beta_1 (\chi_1)$ is to be corrected by including the next order term such that the expression for $\bar \beta_1^2$ included the third order in $\chi_1$ term. The next order term in equation defining $\bar\beta_1 (\chi_1)$ arises since $\bar\beta_1$ should be evaluated at $t=t_1$ or at $\eta=\eta_1=\eta_0-\chi_1$, as follows
\begin{equation}\label{GR3.1.60c}
\bar \beta_1=\frac{da(t)}{dt}\Bigg\vert_{t=t_1}\chi_1=\frac{a'(\eta_1)}{a(\eta_1)}\chi_1
=\frac{a'(\eta_0-\chi_1)}{a(\eta_0-\chi_1)}\chi_1
\end{equation}
Expanding $\bar \beta_1^2$ in series of $\chi_1$ up to the third order yields
\begin{equation}\label{GR3.1.61b}
\bar \beta_1^2=\left(\frac{a'(\eta_0-\chi_1)}{a(\eta_0-\chi_1)}\chi_1\right)^2
=a_0^2 H_0^2 \chi_1^2+2\left(a_0^3 H_0^3-a_0 H_0 x''(\eta_0)\right)\chi_1^3
\end{equation}
where equations (\ref{GR3.1.45}) and (\ref{GR3.1.51}) have been used.
The quantity $x''(\eta_0)$ contained in (\ref{GR3.1.61b}) can be related to the model parameters by exploiting the second Friedmann equation
\begin{equation}\label{GR3.1.62b}
\frac{a''(t)}{a(t)}=-\frac{4\pi G}{3} \rho(t)
\end{equation}
which, with the use of equations (\ref{GR3.1.50}) -- (\ref{GR3.1.53}), can be represented in the form
\begin{equation}\label{GR3.1.63b}
x''(\eta)=\frac{x'(\eta)^2}{x(\eta)}-\frac{1}{2}a_0^2 H_0^2\left(\Omega_M+\Omega_{\Lambda}x(\eta)^3\right)
\end{equation}
Evaluating   (\ref{GR3.1.63b}) with $\Omega_{\Lambda}=0$ at $\eta=\eta_0$ and substituting the result into (\ref{GR3.1.61b}), and then into (\ref{GR3.1.60b}), yields
\begin{equation}\label{GR3.1.64b}
B\left(\bar\beta_1\left(\chi_1\right)\right)=1-\frac{b}{2}\left(\chi_{1m}^2+\Omega_M \chi_{1m}^3\right)
\end{equation}
where $\chi_{1m}$ is defined in (\ref{GR3.1.58c}). Substituting (\ref{GR3.1.64b}) and (\ref{GR3.1.58c}) into (\ref{GR3.1.57}) yields a transcendental equation for $\chi_{1m}$ which can be represented in the form
\begin{eqnarray}\label{GR3.1.65b}
\fl\frac{1-\frac{b}{2}\left(\chi_{1m}^2+\Omega_M \chi_{1m}^3\right)}{1-\Omega_M}\left[\sinh{\left(\frac{1}{2}\sqrt{1-\Omega_M}\chi_{1m}\right)}
-\sqrt{1-\Omega_M}\cosh{\left(\frac{1}{2}\sqrt{1-\Omega_M}\chi_{1m}\right)}\right]^2\nonumber \\
=\frac{1}{1+z}
\end{eqnarray}
where the relation $\Omega_K=1-\Omega_M$ following from equation (\ref{GR3.1.55})  with $\Omega_{\Lambda}=0$ has been  used. Representing the solution of (\ref{GR3.1.65b}) as series in $z$ yields
\begin{equation}\label{GR3.1.66b}
\fl\chi_{1m}(z)=z-\frac{1}{4}\left(\Omega_M+2\left(b +1\right)\right)z^2+\frac{1}{24}\left(8+4\Omega_M+3\Omega_M^2+12b\left(b+1\right)\right)z^3+\cdots
\end{equation}

Having $\chi_1(z)$ defined the luminosity distance can be calculated from (\ref{GR3.1.40}) with $S(\chi_1)$ defined by (\ref{GR3.1.27a}). The expression for  $S(\chi_1)$ can be represented by a single formula which, like equation (\ref{GR3.1.58c}), is valid for all the three cases listed in (\ref{GR3.1.27a}), as follows
\begin{equation}\label{GR3.1.67}
S(\chi_{1})=\frac{1}{a_0 H_0\sqrt{\Omega_K}}\sinh\left(\sqrt{\Omega_K}\chi_{1m}\right)=
\frac{1}{a_0 H_0\sqrt{1-\Omega_M}}\sinh\left(\sqrt{1-\Omega_M}\chi_{1m}\right)
\end{equation}
Then the expression defining $d_L(z)$ is
\begin{equation}\label{GR3.1.68}
d_L(z)=\frac{1}{H_0}\left(1+z\right)\frac{1}{\sqrt{1-\Omega_M}}\sinh\left(\sqrt{1-\Omega_M}\chi_{1m}(z)\right)
\end{equation}
with $\chi_{1m}(z)$ given by (\ref{GR3.1.66b}).

In order to compare the results produced by the model with those, obtained from an analysis of type Ia supernova (SNIa) observations, one needs some fitting formulas for the dependence $d_L(z)$ derived from the observational data. It is now common, in an analysis of the SNIa data, to fit the Hubble diagram of supernovae measurements to the $\Lambda$CDM model (mostly, to  the concordance model) and represent the results as constraints on the model parameters (see, e.g. \cite{Betoule}). Therefore, in what follows, a comparison of the results with the SNIa data is made by comparing the dependence  $d_L(z)$ produced by the present model  with $d_L^{\;c}(z)$ for the concordance model 
with the use of constraints on the parameter $\Omega_M^{\;c}$ from the SNIa data analysis.

\begin{figure}
\centerline{\includegraphics[scale=0.6]{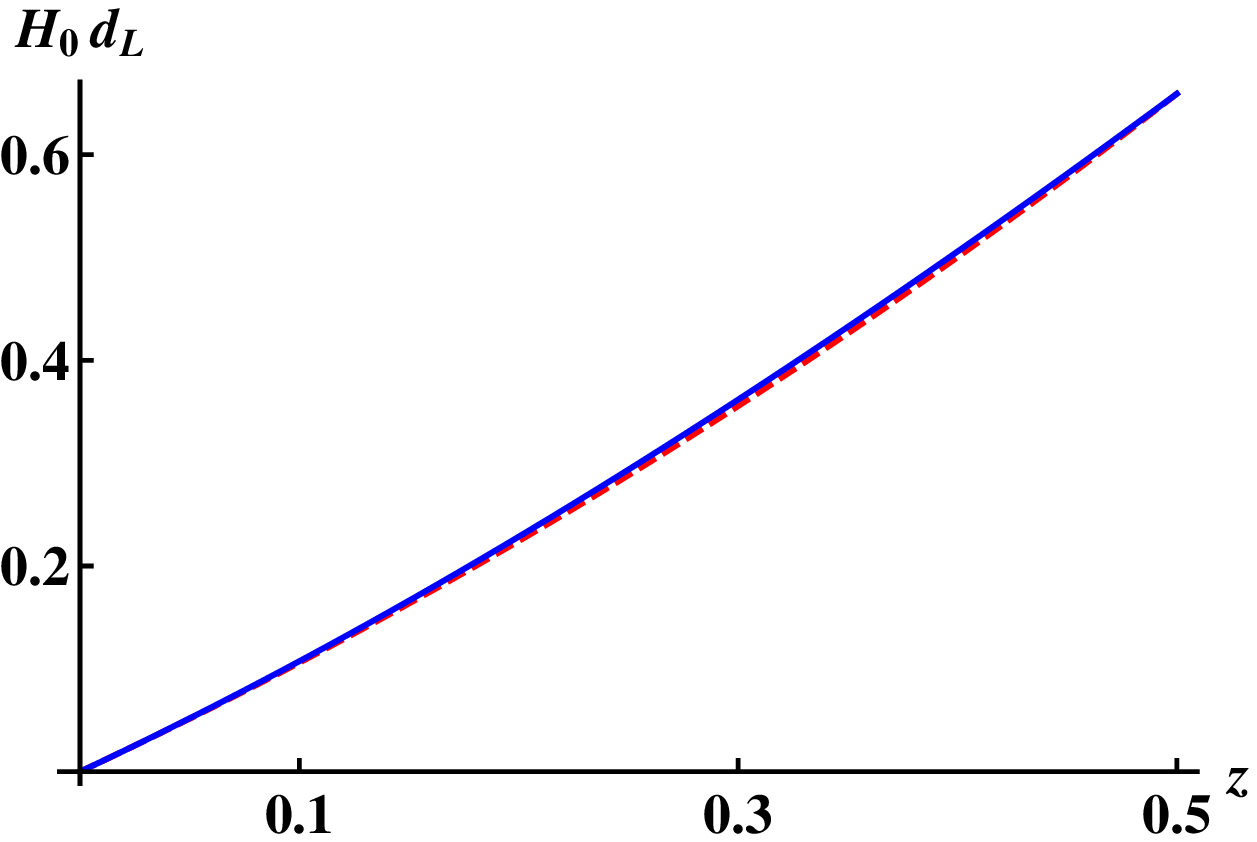} \includegraphics[scale=0.6]{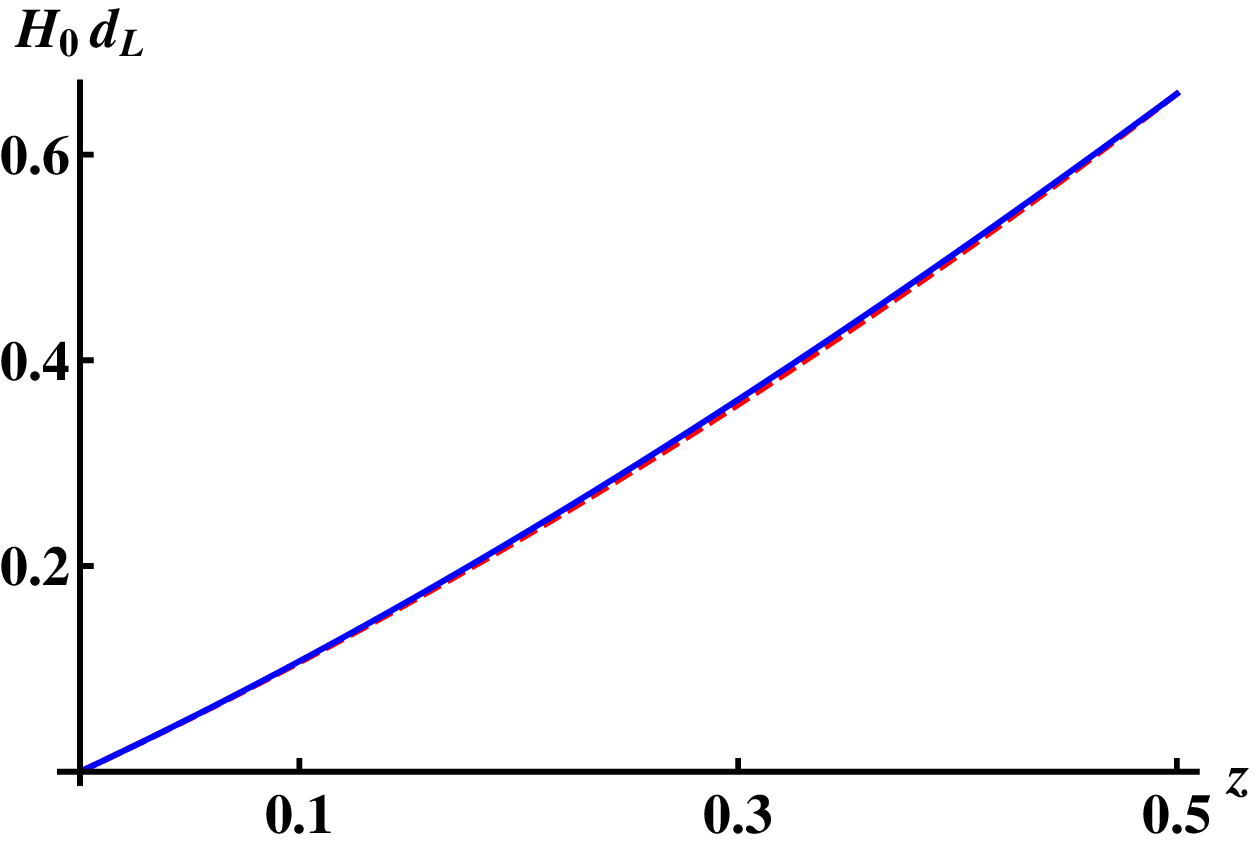}}
\caption{Dependence of the luminosity distance $d_L$ on the red-shift $z$: solid lines for the present model, \textit{left} with  $\Omega_M=1$, $b=-0.672$ (point A  in Fig.\ref{SemPapFig3Alt}) and \textit{right} with $\Omega_M=0.5$, $b=-0.495$ (point B  in Fig.\ref{SemPapFig3Alt}), and a dashed line for the concordance model with $\Omega_M^{\;c\;}=0.31$.}\label{SemPapFig1a}
\end{figure}

\begin{figure}
\centerline{\includegraphics[scale=0.6]{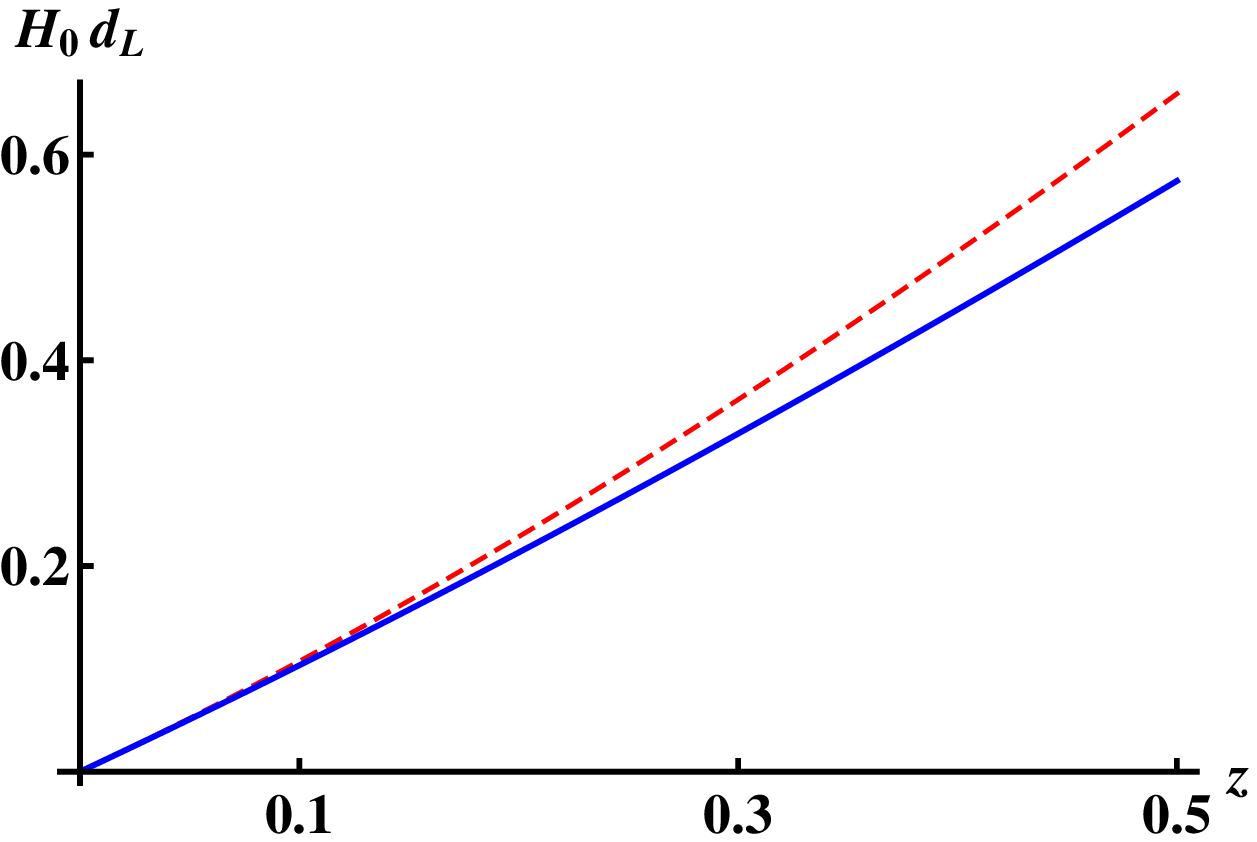} \includegraphics[scale=0.6]{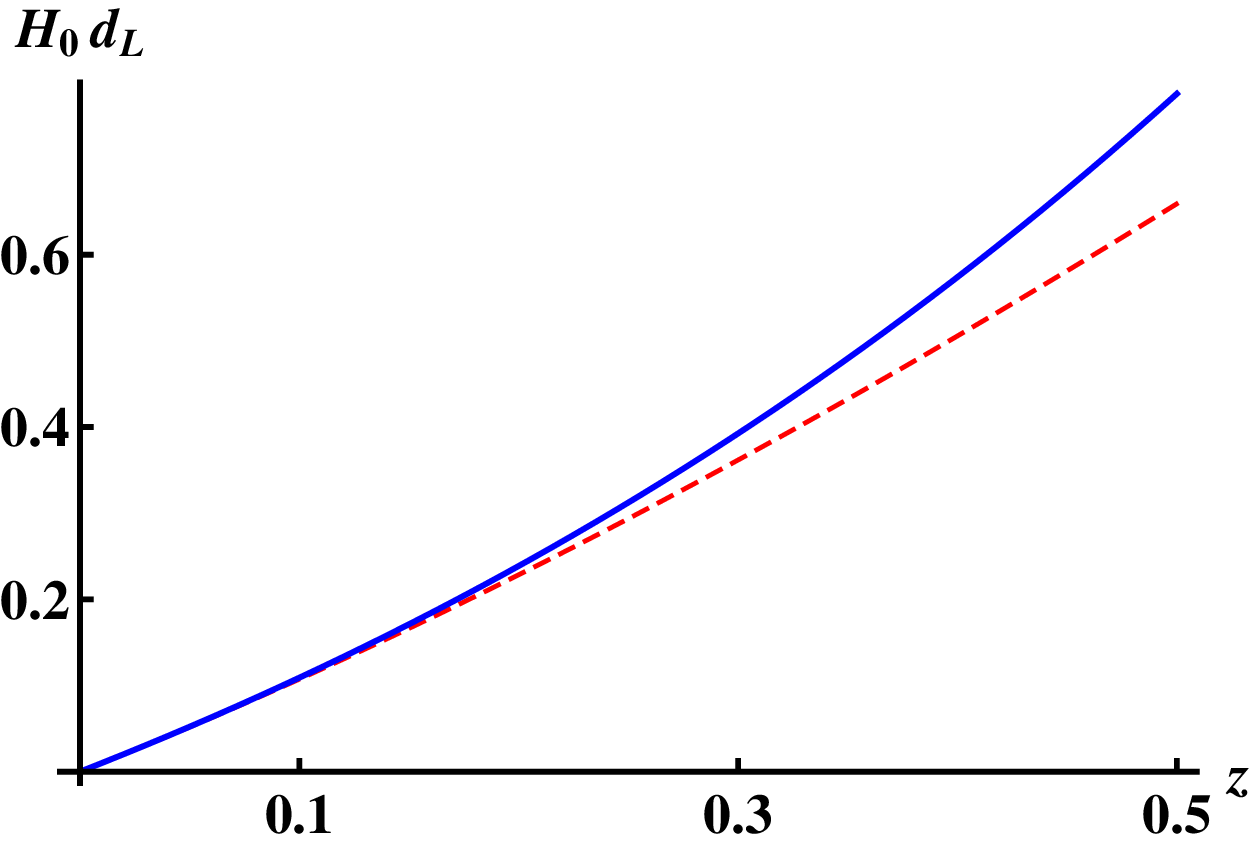}}
\caption{Dependence of the luminosity distance $d_L$ on the red-shift $z$: solid lines for the present model, \textit{left} with  $\Omega_M=1$, $b=-0.2$ (point C in Fig.\ref{SemPapFig3Alt}) and \textit{right} with $\Omega_M=1$, $b=-1.2$ (point D in Fig.\ref{SemPapFig3Alt}) and a dashed line for the concordance model with $\Omega_M^{\;c\;}=0.31$.}\label{SemPapFig1b}
\end{figure}

\begin{figure}
\centerline{\includegraphics[scale=0.9]{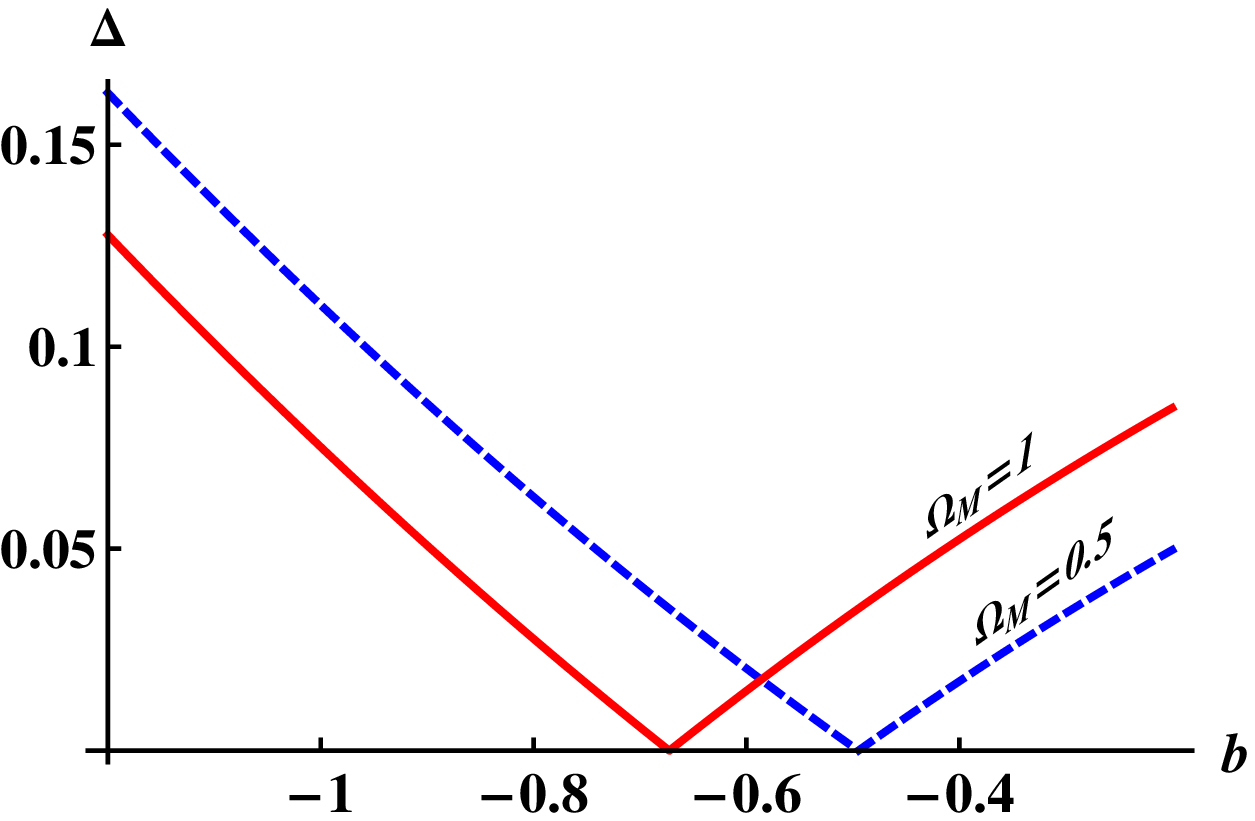}}
\caption{Deviation $\Delta=\vert d_L(z)-d^{\; c \;}_L(z)\vert$ of the luminosity distance $d_L(z)$ produced by the present model from the value $d^{\;c\;}_L(z)$ produced by the concordance model with $\Omega_M^{\;c\;}=0.31$ (calculated for $z=0.5$): a solid line for $\Omega_M=1$ and a dashed line for $\Omega_M=0.5$.}\label{SemPapFig4New}
\end{figure}

A form of $d_L(z)$, defined by equations (\ref{GR3.1.68}) and (\ref{GR3.1.66b}), is governed by two parameters $\Omega_M$ and $b$ while $d_L^{\;c}(z)$, defined by (\ref{GR3.1.70}) and (\ref{GR3.1.69}), depends on a single parameter $\Omega_M^{\;c}$. It is found that, for every value of $\Omega_M^{\;c}$ from the interval, defined by fitting the SNIa data to the concordance model, and for every value of $\Omega_M>0$  the parameter $b$ can be chosen such that the dependence $d_L(z)$ coincided with $d_L^{\;c}(z)$ with a quite high accuracy (were graphically undistinguishable). An example is given in Fig. \ref{SemPapFig1a} while Fig. \ref{SemPapFig1b} shows how it looks if another value of $b$ is chosen. The graphs in Fig. \ref{SemPapFig4New} show the deviation $\Delta=\vert d_L(z)-d^{\; c \;}_L(z)\vert$ as a function of $b$ for two values of $\Omega_M$ and a fiducial value  $\Omega_M^{\;c\;}=0.31$. It is seen that there always exists a value of $b$ for which the deviation is negligible. Note that if another measure of distinction between the graphs is chosen, as for example
\begin{equation}\label{GR3.1.71}
\Delta_1=\frac{\sqrt{\int_0^z{(d_L(\zeta)-d^{\; c}_L(\zeta))^2 d\zeta}}}{\int_0^z{d^{\; c}_L(\zeta)d\zeta}}
\end{equation}
the values of $b$ corresponding to minimal $\Delta_1$ practically coincide with those for minimal $\Delta$.

It is worth clarifying again that the above is intended to be a comparison of the dependence $d_L(z)$ yielded by the present model with that derived from the SNIa observations so that the dependence $d_L^{\;c}(z)$ for the 'concordance' model plays a role of a fitting formula for the SNIa data.


\subsection{Baryon acoustic oscillations}

Baryon acoustic oscillations (BAO) refers to a series of peaks and troughs that are present in the power spectrum
of matter fluctuations due to acoustic waves which propagated in the early universe.
The wavelength of the BAO is related
to the comoving sound horizon at the baryon-drag epoch $r_d$ which
depends on the physical densities of matter.
Measurements of the angular distribution of galaxies yield the quantity
\begin{equation}\label{GR3.1.50b}
\delta\theta(z)=\frac{r_d}{D_M(z)}
\end{equation}
where
$D_M(z)$ is the comoving angular diameter distance related to the physical angular
diameter distance $D_A(z)$ by
\begin{equation}\label{GR3.1.51b}
D_M(z)=\frac{D_A(z) a(t_0)}{a(t_1)}=\frac{a(t_1)S(\chi_1)a(t_0)}{a(t_1)}=a(t_0)S(\chi_1)
\end{equation}
Measurements of the redshift distribution of galaxies yield the quantity $\delta z(z)$ which can be related to the value of $H(z)$, as follows. First, we have
\begin{equation}\label{GR3.1.52b}
r_d=a_0 \delta \chi_1
\end{equation}
Next, according to equation (\ref{GR3.1.55a}), we have
\begin{equation}\label{GR3.1.53b}
\delta\chi_1=\delta\left(\frac{1}{a_0}\int_{x_1}^{1}{\frac{dx}{x'x}}\right)=-\frac{\delta x_1}{a_0 x_1^2 H(x_1)}
\end{equation}

In the standard cosmology, $x_1$ is related to $z$ by equation (\ref{GR3.1.56}) from which it follows that
\begin{equation}\label{GR3.1.54b}
\delta x_1=-\frac {\delta z}{(1+z)^2}=-x_1^2 \delta z
\end{equation}
Combining equations (\ref{GR3.1.52b}) -- (\ref{GR3.1.54b}) yields
\begin{equation}\label{GR3.1.55b}
H^{c\;}(x_1)=\frac{\delta z}{r_d}
\end{equation}

In the present model, $x_1$ is related to $z$ by
\begin{equation}\label{GR3.1.56b}
z+1=\frac{1}{x_1 B\left(\chi_1\left(z\right)\right)}
\end{equation}
so that
\begin{equation}\label{GR3.1.57c}
\delta x_1=-\frac{\left((1+z)B(z)\right)'}{(1+z)^2 B(z)^2}\delta z=-x_1^2 \left((1+z)B(z)\right)'\delta z
\end{equation}
and combining equations (\ref{GR3.1.52b}), (\ref{GR3.1.53b}) and (\ref{GR3.1.57c}) yields
\begin{equation}\label{GR3.1.58b}
H(x_1)=\left((1+z)B(z)\right)'\frac{\delta z}{r_d}
\end{equation}
The function $H(x_1)=x'(t_1)/x(t_1)$ can be expressed through $x(t_1)=x_1$ with the use of the Friedmann equation
(\ref{GR3.1.53}). With the presumption, that in the cosmology based on the relativity with a privileged frame there is no need in introducing dark energy, it is set $\Omega_{\Lambda}=0$ and $H(x_1)$ is obtained from (\ref{GR3.1.53}) in the form
\begin{equation}\label{GR3.1.57d}
H(x_1)=\frac{x'(t_1)}{x(t_1)}=H_0 x_1^{-3/2}\sqrt{\Omega_M+\left(1-\Omega_M\right)x_1}
\end{equation}
where the relation $\Omega_K=1-\Omega_M$ following from equation (\ref{GR3.1.55}) with $\Omega_{\Lambda}=0$ has been used. In order to obtain $H$ as a function of $z$ it is needed to substitute $x_1$ expressed by (\ref{GR3.1.58c}) with a properly defined function $\chi_{1m}(z)$ into (\ref{GR3.1.57d}).
In the papers presenting results of the anisotropic BAO measurements, the quantity $H^{c\;}(z)r_d=\delta z$ is given. Thus, according to equations (\ref{GR3.1.55b}), (\ref{GR3.1.58b}) and (\ref{GR3.1.57d}), in order to check a validity of predictions of the present model one has to compare the quantity
\begin{equation}\label{GR3.1.58d}
H_{cor}(z)=H_0\frac{x_1(z)^{-3/2}\sqrt{\Omega_M+\left(1-\Omega_M\right)x_1(z)}}{\left((1+z)B(z)\right)'}
\end{equation}
with $H^{c\;}(z)$ derived from the BAO data.

The recently released galaxy clustering data set of the Baryon Oscillation
Spectroscopic Survey (BOSS), part of the Sloan Digital Sky Survey III (SDSS III), allowed to obtain
the BAO scales in both transverse and line-of-sight directions.
In \cite{Alam}, the results
of several studies studying that sample with a variety
of methods
are combined into a set of the final consensus constraints that optimally capture all of the information.  The results  are given for three redshift slices centered at redshifts 0.38, 0.51 and 0.61.
The fiducial cosmological model used in that paper is a flat $\Lambda$CDM model with the following parameters: $\Omega_M^{\;c\;}=0.31$ and the Hubble constant
$H_{0,fid}=67.6\; \mathrm{km}\; \mathrm{s}^{-1}\; \mathrm{Mpc}^{-1}$. The sound horizon for this fiducial model is $r_{d,fid}=147.78\; \mathrm{Mpc}$ and constraints are quoted with a scaling factor, e.g., $D_M(z)\times (r_{d,fid}/r_d)$ and $H(z)\times (r_d/r_{d,fid})$. Below the results yielded by the present model are compared with the consensus constraints derived from the BAO data in \cite{Alam}. It is set $H_0=H_{0,fid}=67.6\; \mathrm{km}\; \mathrm{s}^{-1}\; \mathrm{Mpc}^{-1}$ in equations (\ref{GR3.1.58c}) and (\ref{GR3.1.58d}) (in view of the system of units used in the present paper, this value should be divided by $c=299800\; \mathrm{km}\; \mathrm{s}^{-1}$). Using the data of \cite{Alam} we take for $r_d$ the value for a flat $\Lambda$CDM model $r_d=147.78\; \mathrm{Mpc}$, like as in the previous section the SNIa data are identified with their fitting to the flat $\Lambda$CDM model. Since it is a fiducial value of \cite{Alam}, the quantities $D_M(z)\times (r_{d,fid}/r_d)$ and $H(z)\times (r_d/r_{d,fid})$ become equal to $D_M(z)$ and $H(z)$.

\begin{figure}
\centerline{\includegraphics[scale=0.9]{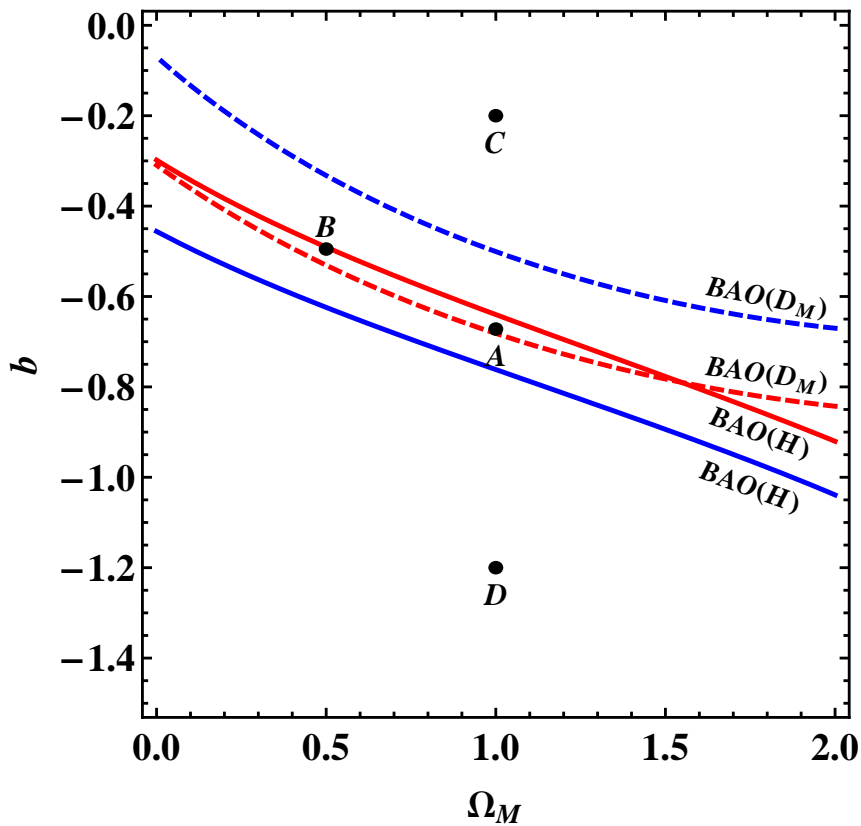}}
\caption{Results of the BAO calculations based on equation (\ref{GR3.1.66b}).
The figure shows contours, on which the values of $H(z)$ and $D_M(z)$
coincide with the values from the consensus constraints for $z=0.38$
inferred from the BAO data (the completed SDSS-III BOSS data set) in  \cite{Alam}: solid contours show the boundaries of the interval obtained using the BAO constraints on $H(z)$: $H^{c\;}(0.38)=81.5 \pm 2.6$; dashed contours show the boundaries of the interval obtained using the BAO constraints on $D_M(z)$: $D_M(0.38)=1518 \pm 31$. The region of overlapping the intervals
is bounded by the contours corresponding to the values $1518 + 31$ (dashed) and $81.5 + 2.6$ (solid). The points "A" and "B" correspond to the values of $b$ (for $\Omega_M=1$ and $\Omega_M=0.5$ respectively) for which a deviation from the SNIa data, as defined in Fig. \ref{SemPapFig4New}, is minimal. The dependence $d_L(z)$ for those points is shown in Fig. \ref{SemPapFig1a} while the graphs $d_L(z)$ for the points "C" and "D" are presented in Fig. \ref{SemPapFig1b}.}\label{SemPapFig3Alt}
\end{figure}

The results for $z=0.38$, based on the third order in $z$ approximate formula (\ref{GR3.1.66b}) for $\chi_{1m}(z)$, are presented in Fig. \ref{SemPapFig3Alt}. The values of $D_M(z)$ are calculated using equations (\ref{GR3.1.51b}) and (\ref{GR3.1.66b}) and the values of $H_{cor}(z)$ are calculated using equations (\ref{GR3.1.58d}), (\ref{GR3.1.58c}) and (\ref{GR3.1.66b}). Boundaries of the regions in the plane $(\Omega_M,b)$, within which the results of the present model are consistent with the constraints on $D_M(z)$ and $H(z)$ from \cite{Alam}, are shown in Fig. \ref{SemPapFig3Alt} by dashed and solid lines respectively. The region of overlapping the intervals corresponds to the values of the model parameters for which the results on $H(z)$ and $D_M(z)$ are consistent both with the BAO data and with each other. It is also consistent with the SNIa data
-- the points corresponding to the values of parameters, for which the deviation $\Delta$ of $d_L(z)$ from the fiducial flat $\Lambda$CDM model is negligible (see Figs \ref{SemPapFig1a} and \ref{SemPapFig4New}), are inside the overlapping region (as, for example, the points "A" and "B").

To show a consistency of the present model results with the constraints from \cite{Alam} for $z=0.51$ and $z=0.61$, another approach, which is not based on the third order in $z$ formula (\ref{GR3.1.66b}) for $\chi_{1m}(z)$, is used. The point is that an accuracy of the third order calculations, acceptable for $z=0.38$, becomes too low for the redshifts 0.51 and 0.61. Calculating $\chi_{1m}(z)$ up to the next (fourth) order in $z$ requires adding the fourth  order in $\chi_{1m}$ term  to the approximation (\ref{GR3.1.61b}) for $\bar\beta_1^2$ and the fourth order in $\bar\beta_1$ term to  the approximation (\ref{GR3.1.60b}) for $B(\bar\beta_1)$. The latter, according to (\ref{B7}), involves one more parameter of the model which makes the analysis more complicated. Instead, an approach based on the assumption  that the model parameters can be chosen such that a value of $d_L(z)$ coincided with the value produced by the flat $\Lambda$CDM model, which is well established in the $z=0.38$ calculations, is applied. Then, according to equations
 (\ref{GR3.1.68}) and (\ref{GR3.1.70}), we have
\begin{equation}\label{GR3.1.59d}
\frac{1}{\sqrt{1-\Omega_M}}\sinh\left(\sqrt{1-\Omega_M}\chi_{1m}(z)\right)=\chi_{1m}^{c}(z)
\end{equation}
from which it follows
\begin{equation}\label{GR3.1.60d}
\chi_{1m}(z)=\frac{1}{\sqrt{1-\Omega_M}}
\sinh^{-1}\left(\sqrt{1-\Omega_M}\chi_{1m}^{c}(z)\right)
\end{equation}
where $\chi_{1m}^{c}(z)$ is given by (\ref{GR3.1.69}). Substituting (\ref{GR3.1.60d}) into the expression (\ref{GR3.1.58c}) for $x_1$ yields
\begin{equation}\label{GR3.1.61d}
x_1(z)=\frac{\left(1-\chi_{1m}^{c}(z)+\sqrt{1+\left(1-\Omega_M\right)\chi_{1m}^{c}(z)^2}\right)^2}
{2\left(1+\sqrt{1+\left(1-\Omega_M\right)\chi_{1m}^{c}(z)^2}\right)}
\end{equation}
Thus, for given $\Omega_M$ and $\Omega_M^{\;c}$ (which enters the expression (\ref{GR3.1.69}) for  $\chi_{1m}^{c}(z)$), $x_1(z)$ can be calculated. Then the factor $B(z)$  can be obtained from (\ref{GR3.1.56b}) and
equation (\ref{GR3.1.58d}) defining $H_{cor}(z)$ takes the form
\begin{equation}\label{GR3.1.62e}
\eqalign {H_{cor}(z)=H_0\frac{x_1(z)^{-3/2}\sqrt{\Omega_M+\left(1-\Omega_M\right)x_1(z)}}{\left(1/x_1(z)\right)'}\cr
=-H_0\frac{x_1(z)^{1/2}\sqrt{\Omega_M+\left(1-\Omega_M\right)x_1(z)}}{x_1'(z)}}
\end{equation}
where $x_1'(z)$ is calculated from (\ref{GR3.1.61d}), as follows
\begin{equation}\label{GR3.1.63e}
x_1'(z)=\left(x_1\left(\chi_{1m}^{c}\right)\right)'\left(\chi_{1m}^{c}\left(z \right)\right)'
\end{equation}
with
\begin{equation}\label{GR3.1.64e}
\fl \left(x_1\left(\chi_{1m}^{c}\right)\right)'=\frac{\left(2-\Omega_M\right)\chi_{1m}^{c}(z)}
{2\sqrt{1+\left(1-\Omega_M\right)\chi_{1m}^{c}(z)^2}}-1,\quad \left(\chi_{1m}^{c}\left(z \right)\right)'=\frac{1}{\sqrt{1-\Omega_M^{\;c}+\Omega_M^{\;c}(1+z)^3}}
\end{equation}
Equations (\ref{GR3.1.61d}) -- (\ref{GR3.1.64e}), with $\chi_{1m}^{c}(z)$ defined by (\ref{GR3.1.69}), provide an exact (not restricted by small $z$) expression for the quantity $H_{cor}(z)$
within the range of parameters where a value of $d_L(z)$ produced by the present model coincides with the value produced by the concordance model.

\begin{figure}
\centerline{\includegraphics[scale=0.9]{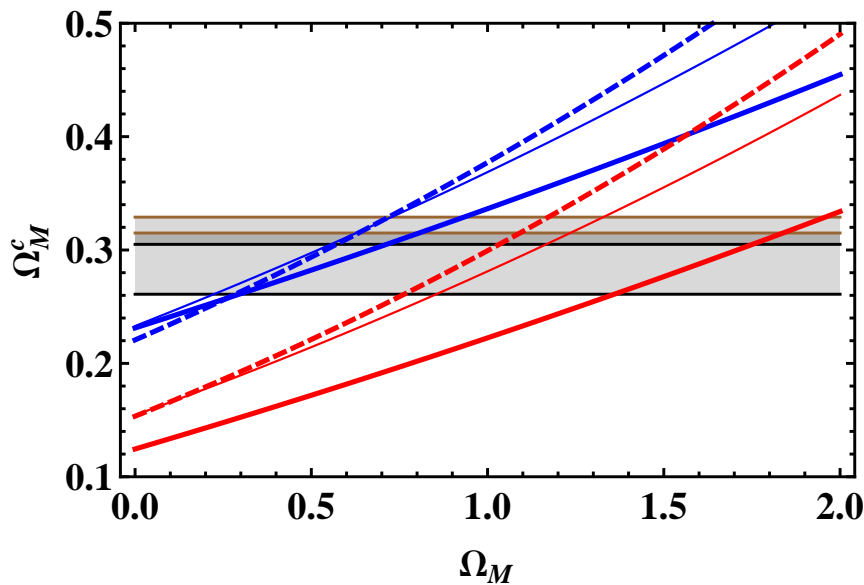}}
\caption{BAO calculations based on equation (\ref{GR3.1.60d}).
 Contours, on which $H_{cor}(z)=H^{c\;}(z)$, with $H^{c\;}(z)$ taken from the constraints  of \cite{Alam}, are shown.  Solid: $H^{c\;}(0.38)=81.5 \pm 2.6$; dashed: $H^{c\;}(0.51)=90.5 \pm 2.7$; dot-dashed: $H^{c\;}(0.61)=97.3 \pm 2.9$. A filled region corresponds to constraints on $\Omega_M^{\;c\;}$ from the SNIa data provided by the SDSS-II and SNLS collaborations \cite{Betoule}: $\Omega_M^{\;c\;}=0.295\pm 0.034$; a darker strip in the middle shows constraints obtained by combining the SNIa and BAO data in \cite{Alam}: $\Omega_M^{\;c\;}=0.310\pm 0.005$.}\label{SemPapFig7New}
\end{figure}

\begin{figure}
\centerline{\includegraphics[scale=0.9]{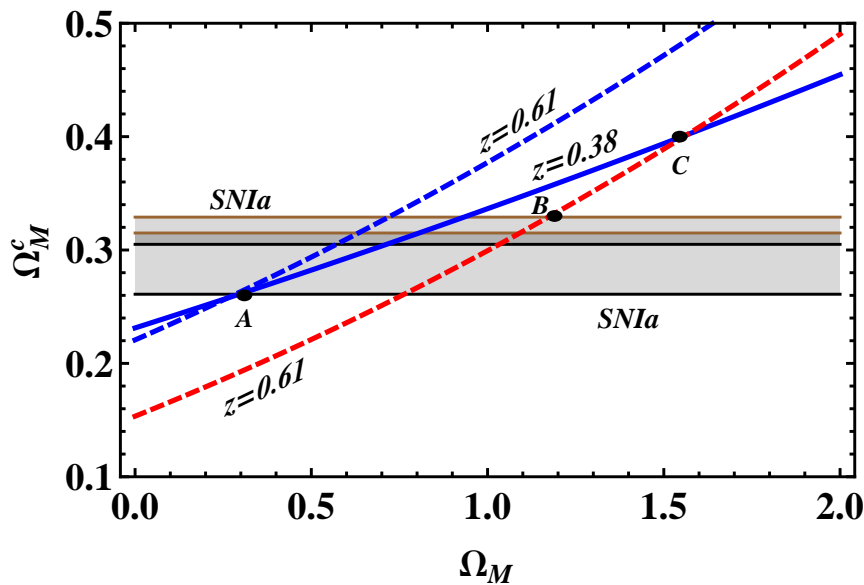}}
\caption{The same as in Fig. \ref{SemPapFig7New} but only with the lines bounding the region wherein the present model results are consistent with the constraints from \cite{Alam}: solid for $z=0.38$ (only the upper line of the two lines in Fig. \ref{SemPapFig7New} is shown) and dashed for $z=0.61$. A filled region shows the same as in Fig. \ref{SemPapFig7New}. Meaning of the points 'A', 'B' and 'C' is discussed in the text.}\label{SemPapFig8New}
\end{figure}

The results are presented in Figs. \ref{SemPapFig7New} and \ref{SemPapFig8New} as contours in the plane $(\Omega_M,\Omega_M^{\;c\;})$ on which $H_{cor}(z)=H^{c\;}(z)$,  with $H^{c\;}(z)=\delta z/r_d$ taken from the consensus constraints of \cite{Alam}. Contours corresponding to constraints on $D_M$ are not shown since, as it can be expected based on equations (\ref{GR3.1.51b}), (\ref{GR3.1.67}) and (\ref{GR3.1.59d}), they do not impose additional restrictions on the parameters.
To demonstrate that the results are consistent both with the constraints of \cite{Alam} for all three redshifts
$z=0.38$, $z=0.51$ and $z=0.61$
and with constraints from the SNIa data provided by the SDSS-II and SNLS collaborations \cite{Betoule}, the latter
are shown as a filled strip in the plane $(\Omega_M,\Omega_M^{\;c\;})$, with a darker strip in the middle showing constraints obtained in \cite{Alam} by combining the SNIa and BAO data.
Fig. \ref{SemPapFig8New} differs from Fig. \ref{SemPapFig7New} in that only the lines bounding the region of parameters, wherein the present model results are consistent with the constraints from \cite{Alam}, are presented. It is seen from  Fig. \ref{SemPapFig8New}, that if only the constraints from the BAO data are taken into account, a restriction on allowed values of $\Omega_M$ is that  they should be smaller than the value corresponding to the point 'C' in Fig. \ref{SemPapFig8New}. (That value $\Omega_M\approx 1.5$  is approximately the same as the value restricting the interval of allowed $\Omega_M$ in Fig. \ref{SemPapFig3Alt}.) However, if the constraints from BAO data of  \cite{Alam} are combined with the constraints on $\Omega_M^{\;c\;}$ from the SNIa data of \cite{Betoule} ($\Omega_M^{\;c\;}=0.295\pm 0.034$), then the interval of allowed $\Omega_M$ becomes narrower being restricted
by the values corresponding to the points 'A' and 'B'. If the constraints on $\Omega_M^{\;c\;}$ from \cite{Alam} ($\Omega_M^{\;c\;}=0.310\pm 0.005$, a darker strip in Fig. \ref{SemPapFig8New}) are used, instead of those of \cite{Betoule}, then the interval is further reduced but still the value of $\Omega_M=1$ belongs to the interval.

Commonly, the BAO observations are considered as confirming the accelerated expansion and imposing constraints on the cosmological parameters in terms of the relative dark energy density $\Omega_{DE}$ and
the parameter $w=p_{DE}/\rho_{DE}$, where $p_{DE}$ and $\rho_{DE}$
are the pressure and density of dark energy respectively. (In the case of $w=-1$, $\Omega_{DE}$ coincides with the above used  $\Omega_{\Lambda}$.) Nevertheless, the only observational data, that may be considered as providing a "direct" evidence for a dark energy, is the Hubble diagram
of distant supernovae. The dark energy arose when  those data were fit to a FLRW cosmology. At present, most of the data provided by a number of independent observations (in particular, the BAO data) are fit to the concordance model, which is the FLRW cosmology with  zero curvature $\Omega_K=0$ and the dark energy obeying an equation of state with $w=-1$ (the flat $\Lambda$CDM model). As it is shown in the previous subsection, the SNIa data can be well fit to the model developed in the present paper - in this model, no acceleration, and correspondingly no dark energy, is needed to explain the data. The results discussed in this subsection show that both the BAO and
SNIa results are consistent with the present model.

\subsection{CMB anisotropies}

The observations of temperature anisotropies in the CMB are commonly considered as providing another independent test for the existence of dark energy.
Like the power spectrum of baryon acoustic oscillations,
the angular power spectrum of CMB temperature ansotropies is dominated by acoustic peaks that arise from gravity-driven sound waves in the photon-baryon fluid in the early universe.
The characteristic angular scale for the location of peaks in the
CMB anisotropy spectrum is given by
\begin{equation}
\theta_M=\frac{r_s(z_{dec})}{D_M(z_{dec})}
\end{equation}
where $r_s(z_{dec})$ is the comoving size of the sound horizon
at the decoupling epoch (at $z_{dec}\approx 1090$)
and $D_M(z_{dec})$ is the comoving angular diameter distance defined by equation (\ref{GR3.1.51b}).
The presence of dark energy
should affect the CMB anisotropies leading to the shift for the positions of acoustic peaks. The most important affect (see, e.g., \cite{Amendola})
is the change of the
position of acoustic peaks due to the modification of the angular diameter
distance coming from a change of $S(\chi_1)$ in equation (\ref{GR3.1.51b}).

It is readily shown that, in the present model,  that change can be attributed to the presence of the factor $B\left(\chi_1\left(z\right)\right)$ in equation (\ref{GR3.1.56b}). The analysis of the previous sections based on the small $z$ approximations cannot be used for calculating $B\left(\chi_1\left(z_{dec}\right)\right)$ and so it cannot be straightforwardly extended to calculating the CMB effects.  (It can be done using some large $z$ approximations but they are less justified and, in addition, involve an undefined parameter needed to be adjusted which, in a sense, is equivalent to adjusting the factor $B$.) 
Nevertheless,  the possibility to fit the CMB data to the model by a proper choice of that factor shows that the data do not contradict the model.

\section{Concluding comments}

Observations of Type Ia supernovae, fitted into
the luminosity distance versus redshift relation of the Robertson-Walker cosmological model of the universe, correspond to the negative deceleration parameter.
The cosmic
acceleration cannot be explained within the context of general relativity if a matter-dominated Friedmann-Robertson-Walker cosmological model of the universe is assumed. Therefore
the dark energy,  a new component of the energy density with strongly negative pressure is introduced.
The concordance of data on the high-redshift supernovae, CMB and BAO with the currently privileged cosmological model (flat $\Lambda$CDM or 'concordance' model) seems to point unambiguously
to the accelerated expansion of the universe and the existence of the dark energy.

Nevertheless, the present's paper analysis shows that those data can be well fit to the model based on the relativity with a privileged frame, in which there is no acceleration and so no dark energy is needed. As it is for the  $\Lambda$CDM model, the data are in concordance with the present model if the values of the model parameters lie within some intervals. As distinct from the $\Lambda$CDM model, fitting the model to the data does not separate the value  $\Omega_K=0$ ($\Omega_M=1$ in the present model) corresponding to the flat universe, although that value is not excluded either. (Possibly, if the model could be applied to the analysis of the CMB data with the same set of model parameters as those used for the analysis of the SNIa and BAO data, some value of $\Omega_M$ were separated.)    It is worthwhile to note that, despite what is frequently claimed, a flatness of the universe is not definitely stated in the modern cosmology. In view of the fact, that there is no direct measurement procedure of the curvature of space independent on the cosmological model assumed, the flatness of the space is the result valid  only within the framework of the $\Lambda$CDM model.

To conclude, one can say that the cosmological model based on the relativity with a privileged frame could provide an alternative to the cosmology with a dark energy.

\ack{The author is grateful to A. Chudaykin for providing useful references.}

\appendix

\section{Factor $B (\bar\beta)$ for an arbitrary relation $k=F(\bar\beta)$}


To represent the integral in  (\ref{GR3.1.2}) as a function of $\bar\beta$, equation (\ref{B3}) and relations $d f(k)=f'(k) dk$,  $f(k)=\bar\beta$ and $k=F(\bar\beta)$ are subsequently used which yields
\begin{equation}\label{B4}
\int_0^k{\frac{p}{\kappa (p)}}d p=\int_0^k\frac{p f'(p)}{1-f^2(p)}d p=\int_0^{\bar\beta}\frac{F(m)}{1-m^2}dm
\end{equation}
Thus, the factor $B (\bar\beta)$ for arbitrary $F(\bar\beta)$ is given by
\begin{equation}\label{B5}
B (\bar\beta)=\exp\left[-\int_0^{\bar\beta}\frac{F(m)}{1-m^2}dm\right]
\end{equation}

The relation (\ref{B5}) can be used, for example, if one wishes to obtain the expansion $d_L(z)$ up to the order of $z^4$. In such a case, the expansion for $z$ should include terms of the order $\chi_1^4$ and, since $\bar\beta_1$ is of the order of $\chi_1$, the factor $B (\bar\beta_1)$ contained in $z$ should include terms of the order of $\bar\beta_1^4$. It requires including the next term into the approximation (\ref{S7}), as follows
\begin{equation}\label{B1}
k=F\left(\bar{\beta}\right)\approx b \bar{\beta}+\alpha \bar{\beta}^3
\end{equation}
With $F(\bar\beta)$ given by (\ref{B1}), the factor $B (\bar\beta)$ defined by (\ref{B5}) becomes
\begin{equation}\label{B6}
B (\bar\beta)=e^{\frac{1}{2}\alpha \bar\beta ^2}\left(1-\bar\beta^2\right)^{\frac{b+\alpha}{2}}
\end{equation}
Equation (\ref{GR3.1.6}) is obtained from (\ref{B6}) as a particular case for $\alpha=0$. Expanding (\ref{B6}) in series up to the order $\bar\beta^4$ yields
\begin{equation}\label{B7}
B (\bar\beta)\approx 1-\frac{1}{2}b\bar\beta^2+\frac{1}{8}\left(\left(b-2\right)b-2\alpha\right)\bar\beta^4
\end{equation}

\section{Gravitational collapse of a dustlike sphere}

In this Appendix, it is considered how the solution of the problem of a spherically symmetric collapse of a 'dust' with negligible pressure (see, e.g., \cite{LL}, \cite{SW1}) is modified with the assumption of the existence of a locally privileged frame. As it follows from the arguments presented in Section 3.1, the solution of the general relativity equations in the general coordinates $(x^0,x^1,x^2,x^3)$ remains valid; the modifications concern only the calculation of physical effects where the proper time and space intervals should be replaced by the 'true' proper time and space intervals. In what follows, the system of units in which $c=1$ is used. With the assumptions of spherical symmetry and negligible pressure of matter, the metric for a cloud of freely falling particles of uniform density being written in the comoving coordinate system and specified using the field equations, takes by a suitable choice of units the form
\begin{equation}\label{GR3.1.15}
ds^2=dt^2-a^2(t)\left[\frac{dr^2}{1-r^2}+r^2 d\Omega\right],\quad d\Omega=d\theta^2+\sin^2\theta d\phi^2
\end{equation}
where the coordinates $r$, $\theta$ and $\phi$ are fixed (time-independent) for a given particle.
Since, in this case, the comoving coordinate system is also synchronous, the variable $t$ is the synchronous proper time at each point of space.  Upon introducing in place of the coordinate $r$ the 'angle' $\chi$ as $r=\sin \chi$, where $0\le \chi\le2\pi$, the metric (\ref{GR3.1.15}) takes the form
\begin{equation}\label{GR3.1.16}
ds^2=dt^2-a^2(t)\left[d\chi^2+\sin^2{\chi}\; d\Omega\right]
\end{equation}
where $\chi$ is a constant for each moving particle.
Solution in a parametric form satisfying the field equations is
\begin{equation}\label{GR3.1.17}
a(t)=a_0 \left(1+\cos\psi\right),\quad t=a_0 \left(\psi +\sin \psi\right)
\end{equation}
where $\psi$   runs from 0 to $\pi$ and $a_0$ is a constant. The density $\rho(t)$ is defined by
\begin{equation}\label{GR3.1.18}
\pi G\rho(t)=\frac{3}{4a_0^2\left(1+\cos\psi\right)^3}
\end{equation}
where $G$ is the gravitational constant.
The solution satisfies the condition that all the particles are at rest at the initial moment $t=\psi=0$ and the collapse moment corresponds to $\psi=\pi$ when all the particles reach the center of the sphere. The constant $a_0$ is determined by the initial conditions: either by the initial radius or by the initial density of the sphere. In the former case, introducing the proper distance $R(\chi,t)$ at time $t$ from the center of the sphere to a co-moving particle with the radial coordinate $\chi$ as $R(\chi,t)=a(t)\chi$
leads to $a_0=R_0/(2 \chi_s)$ where $R_0$ is the radius of the sphere at the moment $t=\psi=0$ and the value $\chi_s$ corresponds to the particles at the sphere surface.

Now, let us assume that the center of the dust sphere is at rest with respect to a privileged frame. To express the solution defined by equations (\ref{GR3.1.16})--(\ref{GR3.1.18}) in terms of 'physical' time $t^{\ast}$ using equation (\ref{GR3.1.12}) the velocity $\bar\beta$ of a given particle with respect to the privileged frame is to be calculated. As it is shown in Section 3.1, modifications due to the presence of a privileged frame
 do not influence the way in which the proper velocity is calculated. However, the proper velocity cannot be calculated exploiting the expression (\ref{GR3.1.14}) for the distance passed by a particle since the comoving coordinates $x^{\alpha}=\{\chi,\theta,\phi\}$ do not change during the particle motion while (\ref{GR3.1.14}) deals with their increments $dx^{\alpha}$. Evidently the trajectories of particles are radial lines so that the particle velocity with respect to the center can be calculated as $\bar v=\partial R(\chi,t)/\partial t$ where $R(\chi,t)$ is the proper distance of the particle with the radial coordinate $\chi$ to the center. With the proper distance defined as $R(\chi,t)=a(t)\chi$, the velocity of a particle with respect to the center (with respect to a privileged frame) is
\begin{equation}\label{GR3.1.19}
\bar \beta=\bar v=\frac{da}{dt}\chi
\end{equation}
(Note that the expression (\ref{GR3.1.19}) is approximate due to an approximate nature of the relation $R(\chi,t)=a(t)\chi$, which is valid only for for 'small' distances, but this approximation is consistent with the approximation used in derivation of (\ref{GR3.1.7})).
Differentiating the relations (\ref{GR3.1.17}) with respect to $t$, while treating $\psi$ as a function of $t$, as follows
\begin{equation}\label{GR3.1.20}
\frac{da}{dt}=-a_0\sin \psi(t)\psi'(t),\quad a_0(1+\cos \psi(t))\psi'(t)=1
\end{equation}
and then eliminating $\psi'(t)$ yields
\begin{equation}\label{GR3.1.21}
\bar\beta=\frac{da}{dt}\chi=-\tan \frac{\psi}{2}\chi,\quad
dt^{\ast}=\left(1-\frac{b}{2}\bar\beta^2\right)dt=\left(1-\frac{b}{2}\chi^2 \tan^2 \frac{\psi}{2}\right)dt
\end{equation}
Relating $dt$ to $d\psi$ by differentiating the second equation of (\ref{GR3.1.17}) and substituting the result into (\ref{GR3.1.21}) yields
\begin{equation}\label{GR3.1.22}
dt^{\ast}=a_0\left(1+\cos{\psi}-\frac{b}{2}\chi^2\left(1-\cos\psi\right)\right)d\psi
\end{equation}
which, upon integration, gives
\begin{equation}\label{GR3.1.23}
t^{\ast}=a_0\left(\psi+\sin{\psi}-\frac{b}{2}\chi^2\left(\psi-\sin\psi\right)\right)\psi
\end{equation}
where the constant of integration has been chosen from the condition $t^{\ast}=0$ when $\psi=0$. This relation allows to express in a parametric form the dependence of the scale factor $a(t)$ on the physical time $t^{\ast}$ and that dependence become different for different particles: $a(t)=\hat a(\chi,t^{\ast})$. In other words, $\psi$, which, according to the second relation of (\ref{GR3.1.17}), was a function of time, becomes a function of $t^{\ast}$ and $\chi$. So the solution of the field equation being expressed in the variables $(t^{\ast},\chi)$ does not correspond to the uniform sphere since the density taken at the same moment of the 'physical' time $t^{\ast}$ becomes a function of $\chi$, as follows
\begin{equation}\label{GR3.1.24}
\rho(t^{\ast},\chi)=\frac{3}{4\pi G a_0^2\left(1+\cos\psi(t^{\ast},\chi)\right)^3}
\end{equation}
Differentiating equation (\ref{GR3.1.24}) with respect to $\chi$ yields
\begin{equation}\label{GR3.1.25}
\frac{\partial \rho(t^{\ast},\chi)}{\partial \chi}=\frac{9 \sin\psi(t^{\ast},\chi)}{4\pi G a_0^2\left(1+\cos\psi(t^{\ast},\chi)\right)^4}\frac{\partial \psi(t^{\ast},\chi)}{\partial \chi}
\end{equation}
It is readily seen that the sign of $\partial \rho(t^{\ast},\chi)/\partial \chi$ coincides with the sign of $\partial \psi(t^{\ast},\chi)/\partial \chi$. The latter can be determined by differentiating equation (\ref{GR3.1.23}) (with $\psi$ replaced by $\psi(t^{\ast},\chi)$) with respect to $\chi$ which yields
\begin{equation}\label{GR3.1.26}
\frac{\partial \psi(t^{\ast},\chi)}{\partial \chi}=\frac{2b \chi\left(\psi(t^{\ast},\chi)-\sin \psi(t^{\ast},\chi)\right)}{2-b \chi^2+\left(2+b \chi^2\right)\cos \psi(t^{\ast},\chi)}
\end{equation}
Analysis of the expression on the right-hand side of (\ref{GR3.1.26}) shows that, in the case of $b<0$, that expression is always negative. In the case of $b>0$, definite conclusions about behavior of $\partial \psi(t^{\ast},\chi)/\partial \chi$ cannot be derived but it is non-monotonic. Thus, in the (more plausible) case of $b<0$, the solution describes a collapsing sphere with the density of the dust
monotonically increasing to the center.

It is evident that, due to the fact that the dependence of the scale factor $a$ on the time $t^{\ast}$ is different for different particles, the particles do nor reach the center simultaneously. However, determining the moment of reaching the center by a given particle from the above solution is not possible since the solution is based on the approximate equation  (\ref{GR3.1.7}) derived under the assumption of not large velocities which is evidently not valid close to singularity.


\begin{thebibliography}{99}


\bibitem{Rob} H.P.~Robertson,
Postulate versus observation in the special theory of relativity.
Rev. Mod. Phys. \textbf{21}, 378 (1949)

\bibitem{MS1} R.~Mansouri and S.U.~Sexl:
A test theory of special relativily: I. Simultaneity and slow clock
synchronization, II. First order tests; III. Second order tests. Gen. Rel. Grav. \textbf{8}, 497--513, 515-524, 809--814 (1977)


\bibitem{Ung1}Ungar, A.A.: Formalism to deal with Reichenbach's
special theory of relativity. Found. Phys. \textbf{21}, 691--726 (1991)

\bibitem{And}Anderson, R., Vetharaniam, I., Stedman, G.E.: Conventionality of synchronisation, gauge dependence
and test theories of relativity. Phys. Rep.  \textbf{295}, 93--180 (1998)

\bibitem{Min}Minguzzi, E.: On the conventionality of simultaneity. Found. Phys. Lett.  \textbf{15}, 153--169 (2002)

\bibitem{Rizzi}Rizzi, G., Ruggiero, M.L., Serafini, A.: Synchronization gauges and the principles of special
relativity. Found. Phys. \textbf{34}, 1835--1887 (2004)



\bibitem{Edw}Edwards, W.F.: Special relativity in anisotropic space. Am. J. Phys. \textbf{31}, 482--489 (1963)

\bibitem{WinII}Winnie, J.A.: Special relativity without one-way velocity assumptions: Part II. Phil. Sci. \textbf{37}, 223--238 (1970)

\bibitem{Ung2}Ungar, A.A.:The Lorentz transformation group of the special theory of relativity without
Einstein's isotropy convention. Phil. Sci. \textbf{53}, 395--402 (1986)

\bibitem{Pauli} Pauli, W.: Theory of Relativity. Pergamon Press, London (1958)

\bibitem{LL} Landau, L.D., Lifshitz, E.M.: The Classical Theory of Fields. Pergamon Press, Oxford (1971)

\bibitem{burde} Burde G.I.: Special relativity kinematics with anisotropic propagation of light and correspondence principle. Found. Phys.,  \textbf{46}, 1573--1597 (2016)


\bibitem{Bate}Bateman, H: The transformation of the electrodynamical euations. Proc. London Math. Soc. \textbf{8},
223--264 (1910)

\bibitem{Cunn}Cunningham, E.:  The principle of relativity in electrodynamics and an extension
 thereof. Proc. London Math. Soc. \textbf{8}, 77--98
(1910)


\bibitem{Fult} Fulton, T., Rohrlich, F., Witten, L.: Conformal invariance in physics. Rev.
Mod. Phys. \textbf{34}, 442--457 (1962)

\bibitem{Kast} H. A. Kastrup, On the advancements of conformal transformations and their
associated symmetries in geometry and theoretical physics. Ann.
Phys. (Berlin) \textbf{17}, 631--690 (2008)

\bibitem{Bogoslov77} See, for example,
Bogoslovsky, G.Yu., Goenner, H.F.: Finslerian spaces possessing local relativistic
symmetry. Gen. Relativ. Gravit. \textbf{31}, 1565--1603 (1999); Bogoslovsky, G.Yu.: Lorentz symmetry violation without violation of relativistic symmetry. Phys. Lett. A \textbf{350}, 5--10 (2006)

\bibitem{Sonego}Sonego, S., Pin, M.: Foundations of anisotropic relativistic mechanics. J. Math. Phys.
\textbf{50}, 042902-1--042902-28  (2009)

\bibitem{Lalan}Lalan, V.: Sur les postulats qui sont \`a la base des cin\'ematiques. Bull. Soc. Math. France \textbf{65}, 83--99 (1937)


\bibitem{Ignatowski} von Ignatowski, W.A.: Einige allgemeine Bemerkungen zum Relativit{\"a}tsprinzip. Phys. Z. \textbf{11}, 972--976 (1910)

\bibitem{FrRo}Frank, Ph., Rothe, H.: Ueber die Transformation der Raumzeitkoordinaten von ruhenden auf bewegte
Systeme. Ann. Phys. \textbf{34}, 825--853 (1911)



\bibitem{Bluman}Bluman, G.W., Kumei, S.: Symmetries and
Differential Equations, Applied Mathematical Sciences, Vol. 81. Springer-Verlag, New York (1989)

\bibitem{Olver} Olver, P.J.: Applications of Lie Groups to
Differential Equations (Graduate Texts in Mathematics: vol 107). Springer, New York (1993)



\bibitem{SW1}S. Weinberg: Gravitation and cosmology: principles and applications of the general theory of relativity. John Wiley \& Sons, Inc. (1972)

\bibitem{SW2}S. Weinberg: Cosmology, Oxford University Press, Oxford (2008)



\bibitem{Betoule} M. Betoule et al., Improved cosmological constraints from a joint analysis of the SDSS-II and SNLS supernova samples. A\&A \textbf{568}, A22 (2014)

\bibitem{Alam}S. Alam et al., The clustering of galaxies in the completed SDSS-III Baryon Oscillation Spectroscopic Survey: cosmological analysis of the DR12. Mon. Not. R. Astron. Soc., \textbf{470},   2617--2652 (2017)

\bibitem{Amendola} L. Amendola and S. Tsujikawa, Dark energy: theory and observations. Cambridge University Press, Cambridge (2010)

\end{thebibliography}
\end{document}